\documentclass[english,aps,prb,twocolumn,superscriptaddress]{revtex4-1}
\usepackage{amssymb}
\usepackage{color}
\usepackage{array}
\usepackage{graphicx}
\usepackage{hyperref}

\hypersetup{colorlinks=false}

\begin{document}

\title{Ballistic charge transport in graphene and light
propagation in periodic dielectric structures with metamaterials: a
comparative study}
\author{Yury P. Bliokh}
\affiliation{CEMS, RIKEN, Saitama, 351-0198, Japan}
\affiliation{Department of Physics, Technion-Israel Institute of Technology, Haifa 32000,
Israel,}

\author{Valentin Freilikher}
\affiliation{CEMS, RIKEN, Saitama, 351-0198, Japan}
\affiliation{Department of Physics, Jack and Pearl Resnick Institute, Bar-Ilan
University, Israel}

\author{Franco Nori}
\affiliation{CEMS, RIKEN, Saitama, 351-0198, Japan}
\affiliation{Department of Physics, University of Michigan, Ann Arbor, Michigan
48109-1040, USA }

\begin{abstract}
We explore the optical properties of periodic layered media containing
left-handed metamaterials. This study is based on several analogies between
the propagation of light in metamaterials and charge transport in graphene.
We derive the conditions for these two problems become equivalent, i.e.,
the equations and the boundary conditions when the corresponding wave
functions coincide. We show that the photonic band-gap structure of a
periodic system built of alternating left- and right-handed dielectric slabs
contains conical singularities similar to the Dirac points in the energy
spectrum of charged quasiparticles in graphene. Such singularities in the
zone structure of the infinite systems give rise to rather unusual
properties of light transport in finite samples. In an insightful numerical
experiment (the propagation of a Gaussian beam through a mixed stack of
normal and meta-dielectrics) we simultaneously demonstrate four Dirac
point-induced anomalies: (i) diffusion-like decay of the intensity at
forbidden frequencies, (ii) focusing and defocussing of the beam, (iii)
absence of the transverse shift of the beam, and (iv) a spatial analogue of
the Zitterbewegung effect. All of these phenomena take place in media with
non-zero average refractive index, and can be tuned by changing either the
geometrical and electromagnetic parameters of the sample,or the frequency and the polarization of light.
\end{abstract}

\maketitle

\affiliation{Department of Physics, Technion-Israel Institute of Technology, Haifa 32000,
Israel,} \affiliation{Advanced Science Institute, RIKEN, Wako-shi, Saitama
351-0198, Japan}

\affiliation{Department of Physics, Jack and Pearl Resnick Institute, Bar-Ilan
University, Israel} \affiliation{Advanced Science Institute, RIKEN,
Wako-shi, Saitama 351-0198, Japan}

\affiliation{Advanced Science Institute, RIKEN, Wako-shi, Saitama 351-0198,
Japan} 
\affiliation{Department of Physics, University of Michigan, Ann Arbor, Michigan
48109-1040, USA }

\affiliation{Department of Physics, Technion-Israel Institute of Technology, Haifa 32000,
Israel,} 
\affiliation{Advanced Science Institute, RIKEN, Wako-shi, Saitama
351-0198, Japan}

\affiliation{Department of Physics, Jack and Pearl Resnick Institute, Bar-Ilan
University, Israel} 
\affiliation{Advanced Science Institute, RIKEN,
Wako-shi, Saitama 351-0198, Japan}

\affiliation{Advanced Science Institute, RIKEN, Wako-shi, Saitama 351-0198,
Japan} 
\affiliation{Department of Physics, University of Michigan, Ann Arbor, Michigan
48109-1040, USA }

\section{Introduction}

Highly unusual properties of monolayers of graphite (graphene) and of
optical media with negative refractive indices (left-handed metamaterials)
had been independently predicted and studied theoretically a long time ago 
\cite{1,2}. At that time, however, these predictions were perceived as
rather intriguing but unrealistic exotica, and remained unnoticed for about
a half-century, until quite recently (and nearly simultaneously) they were
embodied in real materials. This immediately triggered an explosion of
interest and activities, in metamaterials and graphene, both in solid state
physics and optics. Researchers also realized that the most unusual
properties of electron transport in graphene were also peculiar to the
propagation of light in dielectric systems with metamaterials.
Mathematically, this is because, under some (rather general) conditions, the
Maxwell equations for electromagnetic waves in an inhomogeneous dielectric
medium can be reduced to the Dirac equations for charge carriers in graphene
subjected to an external electric potential.

\subsection{Similarities between Maxwell and Dirac equations}

The history of recasting Maxwell equations in alternative, more compact,
spinor forms goes back to the beginning of the past century, and is still in
progress (for a comprehensive historical overview see Refs.~\onlinecite{3, 4}%
, with recent examples in Refs.~\onlinecite{5,6,7,7a}). Therefore it is not
surprising that the similarity between Maxwell and Dirac equations has long
been noticed (according to Ref.~\onlinecite{3}, Majorana discussed it already in 1930).
In the general case of inhomogeneous media, the close analogy between (i) the quantum-mechanical
form of the equations for the Reimann-Silberstain vector fields (linear
combinations of the electromagnetic vectors $\vec{D}$ and $\vec{B}$), and
(ii) the Dirac equation, written in the chiral representation of the Dirac
matrices, was explicitly demonstrated
in Ref.~\onlinecite{3} 

Recently, as graphene became increasingly more popular in solid state
physics, the mathematically established similarity of Maxwell and Dirac
equations took on a new physical significance. Inspired by the very unusual
predictions and discoveries made in graphene, research groups in optics started
endeavors to reproduce the unique transport properties of graphene in
specifically designed dielectric structures. An additional incentive to
these efforts came from the fact that, while the elementary building blocks
of graphene are fixed, modern micro- and nanotechnologies enable
manufacturing periodic dielectric samples with a variety of types and sizes
of unit cells. Moreover, the electrodynamic parameters of photonic crystals
can, in principle, be controlled by external fields, providing unique
opportunities to study condensed matter phenomena in optical ways; for
example, by opening a gap between Dirac cones, as well as breaking and
restoring space-inversion and time-reversal symmetries \cite{5}. Rather
simple electrodynamical analogies furnish physical insights into properties
and applications of graphene such as: the Klein phenomenon \cite{Shytov},
breaking the valley degeneracy \cite{Garcia}, graphene quantum dots \cite%
{Darancet, Rozhkov}, the electronic Goos-Hanchen shift \cite{Chen},
delocalization in one-dimensional disordered systems \cite{8}, etc. 
Furthermore, some exotic phenomena predicted
and discovered in optical systems (like, for example, the ``light wheel'' localized mode \cite{A}, or confined cavity modes and mini-stop bands in broad
periodic photonic waveguides \cite{B,C}) could prompt unique possibilities for
creating new graphene-based devices. In this regard, particularly promising
is the analogy between photonic crystal broad waveguides and zigzag
graphene nanoribbons studied in Ref.~\onlinecite{D}, where it was shown that
the photonic mode coupling also arises in nanoribbons at low energies. The
implementation of the analogy between Dirac electrons and light could be
rewarding for the optical community as well, because it is relatively easy
to create in graphene an inhomogeneous potential pattern with any
distribution of $p-n$ and $n-n$ junctions, while designing a periodic or random
stack of alternating positive-negative dielectric layers is nowadays a
feasible task. Thus, graphene could provide analogue laboratory models in
order to test the optics of metamaterials in a controlled way.

\subsection{Dirac point}

The key feature of graphene, from which all its unique transport properties
stem, is the existence of Dirac cones in the band structure of its energy
spectrum. At first glance, one does not have to work hard to obtain in
optics a double-conical, graphene-like dispersion law: it is inherent in any
plane monochromatic wave propagating in a homogeneous medium, as the
relation between its frequency $\omega $ and the wave number $k$ is given by 
\begin{equation}
\omega ^{2}=c^{2}k^{2}.  \label{eq1}
\end{equation}

\begin{figure}[tbh]
\centering \scalebox{0.2}{\includegraphics{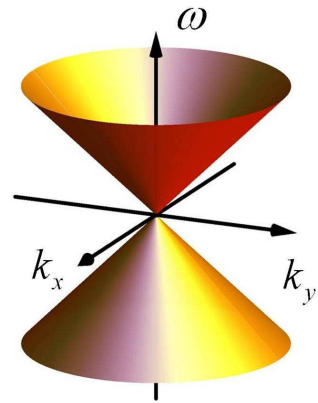}}
\caption{(color online) Surface $\protect\omega (k_{x},k_{y})$ described by
the dispersion equation (1) for electromagnetic waves. The contact of two
cones looks like a Dirac point in graphene; however, in contrast to
graphene, for light both cones correspond to the same field.}
\label{Fig1}
\end{figure}

However, the contact of two cones in Fig.~\ref{Fig1} only looks like a Dirac
point (DP). In fact, of the two cones in Fig.~\ref{Fig1}, only one (for
example, the upper one) is related to a photon, while the second solution of
Eq.~(\ref{eq1}) (lower cone in Fig.~\ref{Fig1}) is redundant. This lower
cone does not carry any additional information and is not related to any
different physical entity like, for example, a hole in graphene or a
positron in the case of relativistic QED. In other words, the photon and
antiphoton are identical \cite{3}. Hence, the challenge in optics is to
create a structure with a real DP in its spectrum, so that different types
of waves would correspond to two different cones (a sort of optical
``particle-antiparticle'' pair). Appropriate for this purpose are photonic
crystals, in which two modes degenerated in a homogeneous space become split
by the periodicity \cite{9}. 

The analytical and numerical studies of two-dimensional periodic structures
(infinite rods embedded in a background medium with a different dielectric
constant) were carried out as early as in 1991, for square \cite{10} and
triangular \cite{11} lattices. Linear singularities (that nowadays are
called Dirac points) are clearly seen in the band structures of both
systems, although they escaped the attention of Refs.~\onlinecite{10,11}
mostly interested in absolute band gaps for different polarizations.
Afterwards, during more than two decades, studies of two-dimensional
photonic crystals were primarily aimed on maximizing the photonic band gap 
\cite{12,15,16} (see also the review Ref.~\onlinecite{17} and references
therein), until the discovery of the unusual transport properties in
graphene, and the potential to reproduce them in optics switched efforts
towards the search and further exploration of photonic structures with
Dirac-cone-like singularities in their transmission spectra \cite{5,7,9,18}.
A number of new optical phenomena arising due to the existence of Dirac
points were predicted and discovered:  diffusion-like $1/L$-dependence of
the pulse intensity on the distance $L$ of propagation inside the photonic
crystal \cite{9,19} (which is unusual for non-random media); oscillatory
motion of a Gaussian beam (optical analogue of the Zitterbewegung effect) \cite%
{20,21,22}; extinction of coherent backscattering \cite{23a,23b,23}; conical
diffraction \cite{24}; as well as the existence of graphene-like and novel
edge states \cite{25, 25a}.

From the above-mentioned publications one can conclude that the existence of
Dirac points in \textit{two-dimensional} periodic structures is a rather
universal phenomenon, in the sense that they appear irrespective of sample
details, such as the shape and dielectric parameters of the ``atoms'' and
its structural symmetry. For example, in the band structures presented in
Refs.~\onlinecite{7,10,11,26}, DPs show up at square, triangular, and
honeycomb lattices. The general criteria for the existence of DP in periodic
dielectric samples were discussed in Ref.~\onlinecite{27}.

The situation in layered periodic media is quite different and less studied.
As we show below, DPs cannot exist in periodically-layered dielectric
structures built of mono-type (i.e., with either all positive or negative
refractive indices) dielectrics, no matter their period, size, and
dielectric contrast between the layers. In Refs.~\onlinecite{27, 34}, a
one-dimensional periodic array of metallic unit cells of a special shape was
considered, which exhibited Dirac points created by the accidental
degeneracy of two modes. To create photonic Dirac cones in one dimension,
both normal and metamaterials should be used. Interestingly enough, a single
DP can exist in a \textit{homogeneous} dispersive meta-medium at a frequency
at which both the dielectric permittivity and the magnetic permeability
approach zero simultaneously \cite{6}. Eigenwaves in \textit{structured},
infinite periodic systems built of alternating normal and left-handed
dielectric layers, and the transmission and reflection from finite samples,
were analyzed in Refs.~\onlinecite{36,37,38}. It was shown that, when the
spatial average of the refractive index over the period was zero, the band
structure consisted of gaps at all frequencies except for a set of isolated
points (discrete modes), for which the optical thicknesses of the adjacent
layers were equal to the same integer number of half-wavelengths. 
New Dirac cones in graphene supelatices created by double-periodic and quasiperiodic electrostatic potentials have been considered in Ref.~\onlinecite{Japan}.

\subsection{Brief summary}

Here we study the transport properties of layered periodic dielectric
systems ``electronically similar'' to graphene, in the sense that they
possess Dirac cones in their photonic band gap structures. In Sec.~II, we
demonstrate, using a simple example, and discuss the similarity and
differences between the Maxwell and Dirac equations as well as between the
corresponding boundary conditions. Section III presents the transmission
through potential barriers created in graphene by applying steplike
electrostatic potentials, in comparison with light propagation in slabs of
normal dielectrics and metamaterials. In Sec.~IV, the photonic band-gap
structures of periodically layered dielectrics are compared with the
structure of the electron energy zones of graphene subjected to a periodic
potential. It is shown that, unlike a two-dimensional photonic crystal, Dirac
cones in a layered periodic medium can exist only when the medium consists
of alternating slabs of left- and right-handed dielectrics (mixed samples).
In Sec. V, we study the propagation of Gaussian beams of light through
periodically-layered \textit{mixed} samples built of alternating slabs with
positive and negative refractive indices, and of beams of charge carriers
through finite graphene superlattices. We demonstrate the anomalous,
diffusion-like dependence of the intensity on the propagation distance, and
an analog of the Zitterbewegung effect in a wide range of parameters. New
unusual transport properties of such samples are predicted. In particular,
it is shown that two contacting Dirac cones manifest
themselves differently: given two beams with frequencies belonging to
different cones, one is focused and another is defocused. The magnitude of
the shift of the focus is independent on the distance of the sample from the
focal plane of the incident beam and is proportional to the width of the
sample. At oblique incidence, a Gaussian beam is not displaced along the
sample even at non-zero values of the mean value of the dielectric constant.

\section{Equations and boundary conditions}

The dynamics of the charge carriers in an external potential $u$ in graphene
is described by a spinor 
\[
\psi =({\psi _{A}},{\psi _{B}})^{T}
\]
whose components are related to two sublattices in the unit cell of the
crystal \cite{39}. In the low-energy limit, near the Dirac point, the
components of this spinor obey the Dirac equations. When the energy $w$ of
the charge carrier is fixed, then the time dependence of the spinor is given
by $\exp (-iwt/\hbar )$, and these equations can be written as 
\begin{eqnarray}
\psi _{A} &=&-\frac{iv_{F}\hbar }{w-u(x,y)}\left( \frac{\partial \psi _{B}}{%
\partial x}-i\frac{\partial \psi _{B}}{\partial y}\right) ,  \nonumber
\label{eq2} \\
\psi _{B} &=&-\frac{iv_{F}\hbar }{w-u(x,y)}\left( \frac{\partial \psi _{A}}{%
\partial x}+i\frac{\partial \psi _{A}}{\partial y}\right) ,
\end{eqnarray}%
where $u(x,y)$ is the electrostatic potential, and $v_{F}$ is the Fermi
velocity.

In order to link this to Maxwell equations, we now consider, as an example,
a TE electromagnetic wave (where the magnetic field has only one non-zero $z$ component) propagating in a homogeneous medium, and introduce two
complex-valued functions 
\begin{equation}  \label{eq3}
\mathcal{E}=E_y-iE_x,\hspace{5mm}\mathcal{H}=ZH_z,
\end{equation}
where $Z=\mu/\varepsilon$ is the medium impedance, $\mu$ and $\varepsilon$
are the medium permeability and permittivity, accordingly. In Ref.~%
\onlinecite{3}, instead of Eq.~(\ref{eq3}), the Reimann-Silberstain vector
wave functions were used to derive a quantum-mechanical, matrix form of the
classical wave equations in the general case of arbitrary electromagnetic
fields propagating in media with space-dependent permittivity and
permeability. The analogy with the relativistic Dirac equations was noted.

It is easy to show that for monochromatic fields $\mathcal{E}$ and $\mathcal{%
H}$ [the time dependence is given by $\exp(-i\omega t)$], the Maxwell
equations yield 
\begin{eqnarray}  \label{eq4}
\mathcal{H}=-{\frac{i}{k_0n}}\left({\frac{\partial\mathcal{E}}{\partial x}}-
i\frac{\partial\mathcal{E}}{\partial y}\right),  \nonumber \\
\mathcal{E}=-{\frac{i}{k_0n}}\left({\frac{\partial\mathcal{H}}{\partial x}}+
i\frac{\partial\mathcal{H}}{\partial y}\right).
\end{eqnarray}
Here, $n$ is the medium refractive index, and $k_0=\omega/c$. It is evident
that after the replacement 
\begin{eqnarray}  \label{eq5}
\mathcal{E}\leftrightarrow\psi_A,  \nonumber \\
\mathcal{H}\leftrightarrow\psi_B,  \nonumber \\
n\omega\leftrightarrow (w-u)/\hbar,  \nonumber \\
c\leftrightarrow v_F\,,
\end{eqnarray}
Eq.~(\ref{eq4}) coincide with Eq.~(\ref{eq2}). Namely, the 2D Maxwell
equations for the complex effective fields (\ref{eq3}) in a homogeneous
medium, and the Dirac equations for the wave functions of the charge carriers in
graphene become identical. The role of the refractive index of the
corresponding effective medium is played by the quantity $n_{\mathrm{eff}%
}=(w-u)/\hbar\omega$. Therefore if, for example, the potential is a
piecewise-constant function of one coordinate, the corresponding graphene
superlattice models a layered dielectric structure \cite{8}. In particular,
a layer, in which the potential $u$ exceeds the energy $w$ of the particle, $%
w-u<0$, is similar to a slab with negative refractive index $n$. This means
that a junction of two regions having opposite signs of $w-u$ is similar to
an interface between left- and right-handed dielectric media with the
refractive indices $n_1$ and $n_2$, if 
\begin{equation}  \label{eq6}
\frac{n_1}{n_2}=\frac{w-u_1}{w-u_2}\,.
\end{equation}
Because of this similarity, a $p-n$ junction can focus Dirac electrons in
graphene \cite{41} in the same way as the focusing of electromagnetic waves
by the boundary between a normal dielectric and a metamaterial \cite{1,42}.
However it is important to realize that, as it follows from Eq.~(\ref{eq6}),
a change of $w$ necessarily implies the corresponding change of the ratio $%
n_1/n_2$. Thus, to model the same (i.e., with fixed values of $u_1$ and $u_2$%
) graphene bi-layer structure, but at different energies, one has to chose
different pairs of dielectrics.

It follows from Eq.~(\ref{eq2}) that the energy spectrum of the charge
carriers in graphene in a homogeneous potential $u=\mathrm{const}$ is linear
near $k=0$, i.e., consists of two cones touching at the
Dirac point (see Fig.~\ref{Fig1}): 
\begin{equation}  \label{eq7}
\left(\frac{w-u}{\hbar v_F}\right)^2=k_x^2+k_y^2.
\end{equation}
After substituting Eq.~(\ref{eq5}), Eq.~(\ref{eq7}) looks exactly like the
dispersion law (\ref{eq1}) for photons. Whilst Eqs.~(2) and (4) are akin,
the similarity between the two problems is not complete. First, unlike Dirac
wavefunctions, the genuine electromagnetic fields $\vec{E}$ and $\vec{H}$
are real, i.e., equal to their complex conjugates. Due to this, positive and
negative frequencies (upper and lower cones in Fig.~\ref{Fig1}) correspond
to the same fields, in contrast to a Dirac spinor, which describes
electrons, when $(w-u)>0$, and holes, when $(w-u)<0$. One further
distinction is that the electric field $\vec{E}$ satisfies the continuity
condition 
\begin{equation}  \label{eq8}
\mathrm{div}\,\vec{E}={\frac{\partial E_x}{\partial x}}+{\frac{\partial E_y}{%
\partial y}}=0\,,
\end{equation}
which is not required for the Dirac wave functions.

Essentially different are also the boundary conditions for the Dirac wave
functions at the interface between two half-spaces with potentials $u_{1}$
and $u_{2}$, 
\begin{equation}
\psi_{1A}=\psi_{2A},\hspace{5mm}\psi_{1B}=\psi_{2B},  \label{eq9}
\end{equation}%
and for the effective electromagnetic fields $\mathcal{E}$ and $\mathcal{H}$
at the boundary between two dielectrics: 
\begin{equation}
Z_2\mathcal{H}_1=Z_1\mathcal{H}_2,\hspace{5mm}\mathrm{Re}\,\mathcal{E}_1=%
\mathrm{Re}\,\mathcal{E}_2.  \label{eq10}
\end{equation}%
Equations~(\ref{eq10}) follow from the boundary conditions for the real
fields, $E_{1y}=E_{2y}$ and $H_{1z}=H_{2z}$ (the boundary is assumed to be
parallel to the $y$ axis ).

Although the charge transport in graphene and the light propagation in
dielectrics are governed by similar equations (\ref{eq2}) and (\ref{eq4})
inside the medium, the boundary conditions (\ref{eq9}) and (\ref{eq10}) are
distinct from each other. This means that, generally speaking, the coupling
between adjacent samples in these two cases is different. This difference is
more conspicuous in the particular case of a monochromatic plane wave, $%
\mathcal{H}\propto\mathcal{E}\propto\exp\left[i\left(k_{x}x+k_{y}y-\omega
t\right)\right] $, when the boundary conditions for the effective fields
take the form: 
\begin{eqnarray}
Z_2\mathcal{H}_1=Z_1\mathcal{H}_2,  \nonumber \\
\mathcal{E}_1\left[1+ik_y/k_{2x}\right]=\mathcal{E}_2\left[1+ik_y/k_{1x}%
\right],  \label{BC}
\end{eqnarray}%
where $k_{jx}=\sqrt{k_0^2n_j^2-k_y^2}$. It is easy to see that Eqs.~(\ref%
{eq9}) and (\ref{BC}) coincide only when $k_{y}=0$ (normal incidence) and $%
Z_{1}=Z_{2}$. That is, in this particular instance, the transmission of
Dirac electrons through a junction is similar to the transmission of light
through an interface between two media with different refractive indices but
equal impedances. In other words, at normal incidence, any junction in
graphene, either $n-n$, $p-p$, or $p-n$, is analogous to a contact between two
perfectly-matched dielectrics or microwave elements. Such an interface is
absolutely transparent to the \textit{normally-incident} radiation and
therefore to the Dirac electrons in graphene as well. This provides a more
intuitive insight into the physics of the Klein paradox (perfect
transmission through a high potential barrier \cite{39, 43}) in graphene
systems.

\section{Transmission through potential barriers and dielectric slabs:
similarities and differences}

To compare the transport properties of Dirac electrons and light in the
general case of oblique incidence, $k_{y}\neq 0$, we first consider the
transmission of particles through a step-like potential barrier, i.e.
through the line $x=0$ separating two domains (1 and 2) of a graphene sheet
with different values $u_{1}$ and $u_{2}$ of the potential. In what follows,
we will consider only one spinor component, say $\psi _{A}\equiv \psi $,
because $\psi _{B}$ could be found using Eq.~(\ref{eq2}). The solutions of
Eqs.~(\ref{eq2}) in both domains can be presented as a linear combination of
plane waves with equal (at the chosen geometry of the system) values of $%
k_{y}$: 
\begin{eqnarray}
\psi _{j}=e^{i\left[ k_{y}y-\left( w-u_{j}\right) t/\hbar \right] }\left\{
\psi _{j}^{(+)}e^{ik_{jx}x}+\psi _{j}^{(-)}e^{-ik_{jx}x}\right\},  \nonumber
\\
j=1,2.\hspace{5mm}  \label{eq11}
\end{eqnarray}%
From now on we consider the range of parameters where there are no total
internal reflections and therefore $\mathrm{Im}k_{x}=0$. From the continuity
of the wave functions at the boundary, Eq.~(\ref{eq9}), it follows (the
incident wave now propagates from medium 1 to medium 2): 
\begin{equation}
\left( 
\begin{array}{c}
\psi _{2}^{(+)} \\ 
\psi _{2}^{(-)}%
\end{array}%
\right) =\hat{M}_{1\rightarrow 2}\left( 
\begin{array}{c}
\psi _{1}^{(+)} \\ 
\psi _{1}^{(-)}%
\end{array}%
\right) ,  \label{eq13}
\end{equation}%
where the transfer matrix $\hat{M}_{1\rightarrow 2}$ for the interface $x=0$
is equal to 
\begin{equation}
\hat{M}_{1\rightarrow 2}={\frac{1}{2\cos \theta _{2}}}\left\vert \left\vert 
\begin{array}{cc}
g_{1\rightarrow 2}^{(+)} & g_{1\rightarrow 2}^{(-)}\vspace{2mm} \\ 
{g_{1\rightarrow 2}^{(-)}}^{\ast } & {g_{1\rightarrow 2}^{(+)}}^{\ast }%
\end{array}%
\right\vert \right\vert  \label{eq14}
\end{equation}%
and the matrix elements are 
\begin{equation}
g_{1\rightarrow 2}^{(\pm )}=e^{-i\theta _{2}}\pm s_{1}s_{2}e^{\pm i\theta
_{1}}.  \label{eq15}
\end{equation}%
Here $s_{j}=\mathrm{sgn}(w-u_{j})$, $\theta _{1}$ and $\theta _{2}$ are the
angles of incidence and refraction, respectively, while $\theta _{j}=\arctan
(k_{y}/k_{jx})$.

We determine the transmission $T$ and reflection $R$ coefficients as the
ratios of the normal-to-the-boundary components of the densities of the
transmitted, $\vec{J}_{2}^{(+)}$, and reflected, $\vec{J}_{1}^{(-)}$,
currents divided by the normal component of the incident current density ${J}%
_{1x}^{(+)}$: 
\begin{equation}
T={J}_{2x}^{(+)}/{J}_{1x}^{(+)},\hspace{5mm}R={J}_{1x}^{(-)}/{J}_{1x}^{(+)},
\label{eq16}
\end{equation}%
where 
\begin{equation}
J_{jx}^{(\pm )}=\pm 2s_{j}\cos \theta _{j}\left\vert \psi _{j}^{(\pm
)}\right\vert ^{2}.  \label{eq17}
\end{equation}

From Eqs.~(\ref{eq16}), (\ref{eq17}) the following formulas can be obtained
(see, e.g., Refs.~\onlinecite{Bliokh, Allian}): 
\begin{eqnarray}  \label{eq18}
T=\frac{2\cos\theta_1\cos\theta_2}{1+\cos\left(s_1\theta_1+s_2\theta_2\right)%
}  \nonumber \\
R=\frac{1-\cos\left(s_1\theta_1-s_2\theta_2\right)}{1+\cos\left(s_1%
\theta_1+s_2\theta_2\right)},
\end{eqnarray}
where the angles of incidence $\theta_1$ and refraction $\theta_2$ are
connected by the relation 
\begin{equation}  \label{eq19}
\sin {\theta _2}=\frac{{w-{u_1}}}{{w-{u_2}}}\sin{\theta_1}.
\end{equation}
The same signs of $s_1$ and $s_2$ correspond to $n-n$ (or $p-p$) junctions,
while for a $n-p$ (or $p-n$) junction $s_1=-s_2$. It is easy to see that at
normal incidence ($\theta_1=\theta_2=0$) the potential barrier of any height
is absolutely transparent at any energy (the Klein tunneling effect).

In the case of light propagating through the interface between two
dielectrics, whose parameters are $\varepsilon_1$, $\mu_1$ and $%
\varepsilon_2 $, $\mu_2$, the transfer matrix 
\begin{equation}  \label{eq20}
\hat{\mathcal{M}}_{1\rightarrow 2}={\frac{1}{2\cos \theta _{2}}}\left\| 
\begin{array}{cc}
\mathcal{G}_{1\rightarrow 2}^{(+)} & \mathcal{G}_{1\rightarrow 2}^{(-)} \\ 
{\mathcal{G}_{1\rightarrow 2}^{(-)}}^{\ast } & {\mathcal{G}_{1\rightarrow
2}^{(+)}}^{\ast }%
\end{array}%
\right\|
\end{equation}
connects the amplitudes of leftward and rightward propagating waves at both
sides, and has the same form as Eq.~(\ref{eq14}), with the matrix elements $%
g_{2 \rightarrow 1}^{\left(\pm\right)}$ replaced by 
\begin{equation}  \label{eq21}
\mathcal{G}_{1\rightarrow 2}^{\left( \pm \right)} = \cos {\theta _2} \pm {s_1%
}{s_2}\cos {\theta _1}\frac{{Z_2}}{{Z_1}}
\end{equation}
for TE waves, and by 
\begin{equation}  \label{eq22}
\mathcal{G}_{1\rightarrow 2}^{\left( \pm \right)} = \cos {\theta _2}\frac{{%
Z_2}}{{Z_1}} \pm {s_1}{s_2}\cos {\theta _1}
\end{equation}
for TM radiation. In Eqs.~(\ref{eq21}) and (\ref{eq22}), $s_j=\mathrm{sgn}\,n_j$%
, $n_j=\pm\sqrt{\varepsilon_j\mu_j}$, and $Z_j=\sqrt{\mu_j/\varepsilon_j}$. For
left-handed dielectrics ($\varepsilon<0$, $\mu<0$), the refractive index $n$
is negative.

When $Z_1=Z_2$, the expressions (\ref{eq21}) and (\ref{eq22}) are identical
and are related to the matrix elements (\ref{eq15}) of the corresponding
transfer matrix in graphene, Eq.~(\ref{eq13}), as 
\begin{equation}  \label{eq23}
\mathcal{G}_{1\rightarrow 2}^{(\pm)}=\mathrm{Re}\, g_{1\rightarrow
2}^{(\pm)}.
\end{equation}
This relation is a consequence of the abovementioned difference between the
solutions of the Dirac and Maxwell equations: the former are complex-valued
functions, while the electromagnetic fields are real.

The light transmission and reflection coefficients are determined as the
ratios of the normal components of the transmitted (for $T$) and reflected
(for $R$) energy fluxes divided by the normal component of the incident
energy flux. When $Z_1=Z_2$ (the situation most favorable for the analogy
between Dirac electrons and light) $T$ and $R$ take the forms: 
\begin{eqnarray}  \label{eq24}
T=\frac{4\cos\theta_1\cos\theta_2}{\left(\cos\theta_1+\cos\theta_2\right)^2},
\nonumber \\
R=\frac{\left(\cos\theta_1-\cos\theta_2\right)^2}{\left(\cos\theta_1+\cos%
\theta_2\right)^2}.
\end{eqnarray}
One can see that even in the particular case, $Z_1=Z_2$, Eqs.~(\ref%
{eq18}) and (\ref{eq24}) are, generally speaking, different, and coincide
only when $\theta_1=\theta_2=0$, i.e., when the boundary conditions Eq.~(\ref%
{BC}) are equivalent. This means that in spite of the identity of 
Eqs.~(\ref{eq2}) and (\ref{eq4}), the analogy between the transport of
Dirac electrons in graphene and electromagnetic radiation in dielectrics
should not be extended too far. Due to the differences in the boundary
conditions, the analogy holds only for normal incidence on the interface
between two perfectly matched media.

The transmission of Dirac electrons trough a potential barrier of finite
width $d$, and of an electromagnetic wave through a dielectric slab of the
same width are described by similar matrices of the form 
\begin{equation}  \label{eq25}
\hat{B}=\hat{A}_{2\rightarrow 1}\hat{S}_{d}\hat{A}_{1\rightarrow 2},
\end{equation}
where the indices 1 and 2 now correspond, respectively, to the outside and
inside of the barrier (slab). The matrix $\hat{A}$ is equal to $\hat{M}$, as
in Eq.~(\ref{eq14}) for graphene, and $\hat{A}=\hat{\mathcal{M}}$, as in
Eq.~(\ref{eq20}) for dielectrics. Obviously, $\hat{A}_{2\rightarrow 1}\hat{A}%
_{1\rightarrow 2}=\hat{I}$, where $\hat{I}$ is a unit matrix. The diagonal
matrix $\hat{S_d}=\mathrm{diag}\left(e^{i\varphi},e^{-i\varphi}\right)$,
with $\varphi=k_{2x}d$ describes the propagation inside the barrier
(dielectric slab). From Eq.~(\ref{eq25}) it follows that under the condition 
\begin{equation}  \label{eq26}
k_{2x}d=m\pi,\hspace{5mm} m=1,2,3,\ldots,
\end{equation}
$\hat{B}=(-1)^m\hat{I}$. This means that either a potential barrier or a
layer of dielectric is transparent if its width is equal to an integer
number of half-wavelengths. This is a general property of both, Dirac and
Maxwell equations, which is independent of the ratio between the impedances,
and its physical nature has nothing to do with the Klein tunneling.

\section{Periodically-stripped graphene superlattices and
photonic structures}

In this section, we compare the structure of the electron energy zones of
graphene subject to a periodic electrostatic potential, with the photonic
band gap structures of periodically-layered dielectrics. To do this we
assume that the potential in graphene takes the two values $u_1$ and $u_2$
in alternating areas of  widths $d_1$ and $d_2$, and the
corresponding dielectric sample is built of alternating layers of the same
thicknesses, $d_1$ and $d_2$, with refractive indices $n_1$ and $n_2$, and
impedances $Z_1$ and $Z_2$, respectively. In both cases, the propagation in
the layers 1 and 2 is described by the matrices $\hat{S}_j=\mathrm{diag}%
\left(e^{i\varphi_j},e^{-i\varphi_j}\right)$, $j=1,2$, where $%
\varphi_{j}=d_{j}k_{jx}$. Assuming that the layers are parallel to the $y$%
-axis, the transformation matrix $\hat{P}$ on the period $D=d_1+d_2$ is
defined by $\psi(x+D)=\hat{P}\psi(x)$, and is equal to 
\begin{equation}  \label{eq27}
\hat{P}=\hat{S}_1\hat{M}_{2\rightarrow 1}\hat{S}_2\hat{M}_{1\rightarrow 2}.
\end{equation}
for graphene, and 
\begin{equation}  \label{eq28}
\hat{\mathcal{P}}=\hat{S}_1\hat{\mathcal{M}}_{2\rightarrow 1}\hat{S}_2\hat{%
\mathcal{M}}_{1\rightarrow 2}.
\end{equation}
for dielectrics.

The eigenvalues $\lambda =\exp (ik_{\parallel }D)$ ($k_{\parallel }$ is the
Bloch wavenumber along the $x$-axis) of the matrix $\hat{P}$ ($\hat{\mathcal{%
P}}$) depend on the energy (frequency) and on the tangent component $k_{y}$
of the wavevector. Both periodic structures are transparent if 
\begin{equation}
|\lambda |=1.  \label{eq29}
\end{equation}%
In other words, Eq.~(\ref{eq29}) determines the transparency zones: the
ranges of the energies (frequencies) and wave numbers $k_{y}$ for which the
longitudinal wavenumber $k_{\parallel }$ is real \cite{45}. It can be shown
that 
\begin{equation}
\lambda =F\pm \sqrt{F^{2}-1},  \label{eq30}
\end{equation}%
where 
\begin{eqnarray}
F(w,k_{y})= \cos \varphi _{1}\cos \varphi _{2}  \nonumber \\
+\left( \tan \theta _{1}\tan \theta _{2}-{\frac{s_{1}s_{2}}{\cos \theta
_{1}\cos \theta _{2}}}\right) \sin \varphi _{1}\sin \varphi _{2}
\label{eq31}
\end{eqnarray}%
for Dirac electrons in graphene and 
\begin{eqnarray}
F(\omega ,k_{y})=\cos \varphi _{1}\cos \varphi _{2}  \nonumber \\
-{\frac{s_1s_2}{2}}\left(\frac{Z_1\cos\theta_1}{Z_2\cos\theta_2}+\frac{%
Z_2\cos\theta_2}{Z_1\cos\theta_1}\right) \sin \varphi _{1}\sin \varphi _{2}.
\label{eq32}
\end{eqnarray}%
for electromagnetic waves in layered dielectrics. In both cases, the
eigenmodes obey the dispersion equation: 
\begin{equation}
\cos (k_{\parallel }D)=F.  \label{eq32a}
\end{equation}

It follows from Eqs.~(\ref{eq29}) and (\ref{eq30}) that a periodic structure
is transparent for the points in the plane $(k_y,w)$ [or $(k_y,\omega)$] for
which the inequality $|F|\leq 1$ holds. One the other hand, in each of these
two planes, the conditions 
\begin{eqnarray}  \label{eq33}
\varphi_1=p\pi,\hspace{5mm}p=1,2,3,\ldots\,,  \nonumber \\
\varphi_2=q\pi,\hspace{5mm}q=1,2,3,\ldots\,
\end{eqnarray}
($p$ and $q$ are positive integer numbers) determine two sets of curves,
where $\hat{P}$ and $\hat{\mathcal{P}}$ are equal to $(-1)^{p+q}\hat{I}$.
This means that those types of curves belong to transparency zones. It is
apparent that since at the crossings of these lines, the eigenvalues of the
matrices $\hat{P}$ and $\hat{\mathcal{P}}$ are equal to $\pm 1$, such
singular crossing points lie at the edges of the zones and represent their
points of contact, known as  ``conical intersection,'' or ``diabolic points'' (see, e.g., Ref.~%
\onlinecite{Berry}).

The coordinates, $k_{yt}$ and $k_{0t}=\omega_t/c$, of these points in the
plane $(k_y,k_0)$ can be found from Eqs.~(\ref{eq33}): 
\begin{eqnarray}  \label{eq34}
k_{0t}^2=\left[\left(p\pi/d_1\right)^2-\left(q\pi/d_2\right)^2\right]%
/\left(n_1^2-n_2^2\right),  \nonumber \\
k_{yt}^2=\left[n_2^2\left(p\pi/d_1\right)^2-n_1^2\left(q\pi/d_2\right)^2%
\right]/\left(n_1^2-n_2^2\right).
\end{eqnarray}
In the case of graphene, the coordinates of the diabolic points in the plane 
$(w,k_y)$ are calculated in the same way. In the vicinity of the points defined by Eq.~(%
\ref{eq33}), the phases $\varphi_1$ and $\varphi_2$ can be written as 
\begin{equation}  \label{eq35}
\varphi_1=p\pi+\delta\varphi_1,\hspace{5mm}\varphi_2=q\pi+\delta\varphi_2,
\end{equation}
where $|\delta\varphi_j|\ll 1$. Substituting Eqs.~(\ref{eq35}) into Eq.~(\ref%
{eq32}) yields: 
\begin{eqnarray}  \label{eq36}
F\simeq(-1)^{p+q}\left[1-{\frac{1}{2}}\left(\delta\varphi_1^2+\delta%
\varphi_2^2+2a\delta\varphi_1\delta\varphi_2\right)\right]  \nonumber \\
\equiv \left[1-{\frac{1}{2}}f(\delta\varphi_1,\delta\varphi_2)\right]
\end{eqnarray}
with 
\begin{eqnarray*}
a={\frac{s_1s_2}{2}}\left(\frac{Z_1\cos\theta_1}{Z_2\cos\theta_2}+\frac{%
Z_2\cos\theta_2}{Z_1\cos\theta_1}\right) \\
\equiv{\frac{1}{2}}\left({\frac{k_{1x}\varepsilon_2}{k_{2x}\varepsilon_1}}+{%
\frac{k_{2x}\varepsilon_1}{k_{1x}\varepsilon_2}}\right),
\end{eqnarray*}
where the components $k_{1x}$ and $k_{2x}$ are taken at the point $%
(k_{yt},k_{0t})$. In the plane $(\delta k_y,\delta k_0)$, where $\delta k_0$
and $\delta k_y$ are small deviations of $k_0$ and $k_y$ from their values
given by Eq.~(\ref{eq34}), the points for which the quadratic form $%
f(\delta\varphi_1,\delta\varphi_2)$ (\ref{eq36}) is positive, constitute a
transparency zone, while the points with $f<0$ correspond to a gap in the
photonic spectrum.

The relation between $\delta \varphi _{j}$, $\delta k_{y}$, and $\delta k_{0}
$ follows from the formula: 
\begin{equation}
\varphi _{j}=d_{j}\sqrt{n_{j}^{2}(k_{0})k_{0}^{2}-k_{y}^{2}},  \label{eq37}
\end{equation}%
where $n_{j}(k_{0})$ is the refraction index of the $j$-th layer. An
analogue formula for graphene has the form: 
\begin{equation}  \label{eqG1}
\varphi_j=d_j\sqrt{\left[(w-u_j)/\hbar v_F\right]^2-k_y^2},
\end{equation}

\begin{figure}[tbh]
\centering \scalebox{0.35}{\includegraphics{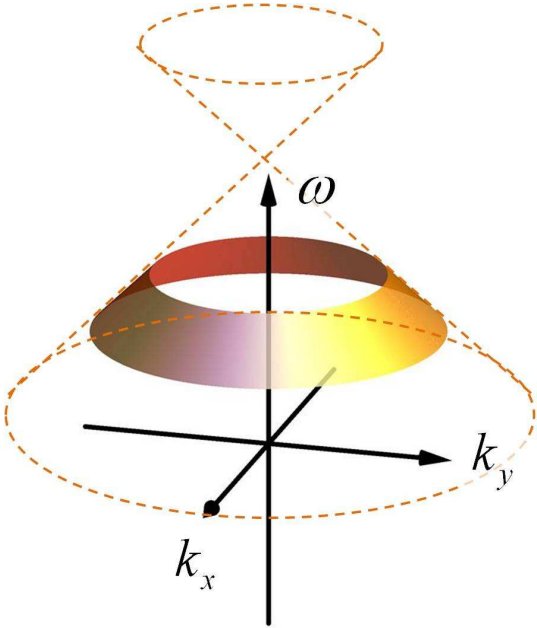}}
\caption{(color online) Surface $\protect\omega(k_x,k_y)$ in a small
vicinity of a certain frequency for left-handed media.}
\label{Fig2}
\end{figure}

While in conventional dielectrics, the dispersion can be ignored, if it is
weak enough, in left-handed materials it must always be taken into account.
Indeed, the surface $\omega (k)$, where $k=\sqrt{k_{x}^{2}+k_{y}^{2}}$, for
normal dielectrics, is a cone similar to the one presented in Fig.~\ref{Fig1}%
; i.e., $\omega (k)=n^{-1}ck$, with $n=\mathrm{const}>0$. For left-handed
metamaterials $n<0$, and the group velocity $v_{g}=(d\omega /dk)$ is
negative, $v_{g}<0$, i.e., is antiparallel to the phase velocity $\omega/k$.
The surface $\omega (k)$, in a small vicinity of a certain frequency $\omega
_{0}$, has the form depicted in Fig.~\ref{Fig2}. Although locally it looks
like a part of the lower cone in Fig.~\ref{Fig1}, $\omega(k)\neq n^{-1}ck$,
with $n=\mathrm{const}$ on this surface, because the contact
point of the cones is shifted from the origin. 

\begin{figure}[tbh]
\centering \scalebox{0.35}{\includegraphics{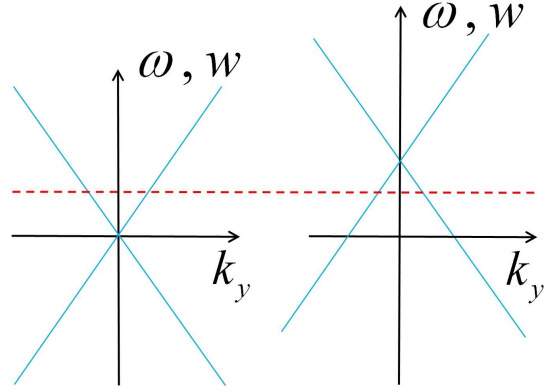}}
\caption{(color online) The dispersion characteristics $\protect\omega(k)$,
shown as blue lines, of two adjoined right-handed (left panel) and
left-handed (right panel) dielectric layers is identical to the energetic
spectrum $w(k_y)$ of $n-p$ junction in graphene. The horizontal dashed red line
represents the wave frequency, or the quasi-particle energy.  The cones' apexes correspond to the zones touching points.}
\label{TwoLayersDisp}
\end{figure}

Although the absolute values of the transmission and reflection coefficients
in Eqs.~(\ref{eq18}) and (\ref{eq24}) are different, the dispersion
characteristics of two adjoining right- and left-handed dielectric layers,
and the energetic spectrum diagrams of $n-p$ junction in graphene are
identical, as it is shown schematically in Fig.~\ref{TwoLayersDisp}. As a
consequence of this identity, the periodic dielectric structure and graphene
superlattice formed by a periodic external potential possess the same unique
transport properties. 

\subsection{Photonic structure}

From Eq.~(\ref{eq37}) we obtain: 
\begin{equation}
\delta \varphi _{j}=\frac{d_{j}}{k_{jx}}\left( \frac{ck_{jt}}{v_{jg}}\delta
k_{0}-k_{yt}\delta k_{y}\right) ,  \label{eq38}
\end{equation}%
where $v_{jg}$ is the wave group velocity in the $j$-th layer at the
frequency $\omega _{t}=ck_{0t}$. The group velocity in Eq.~(\ref{eq38}) is
positive for normal dielectric layers, and negative for layers of
left-handed metamaterials.

\begin{figure}[tbh]
\centering \scalebox{0.6}{\includegraphics{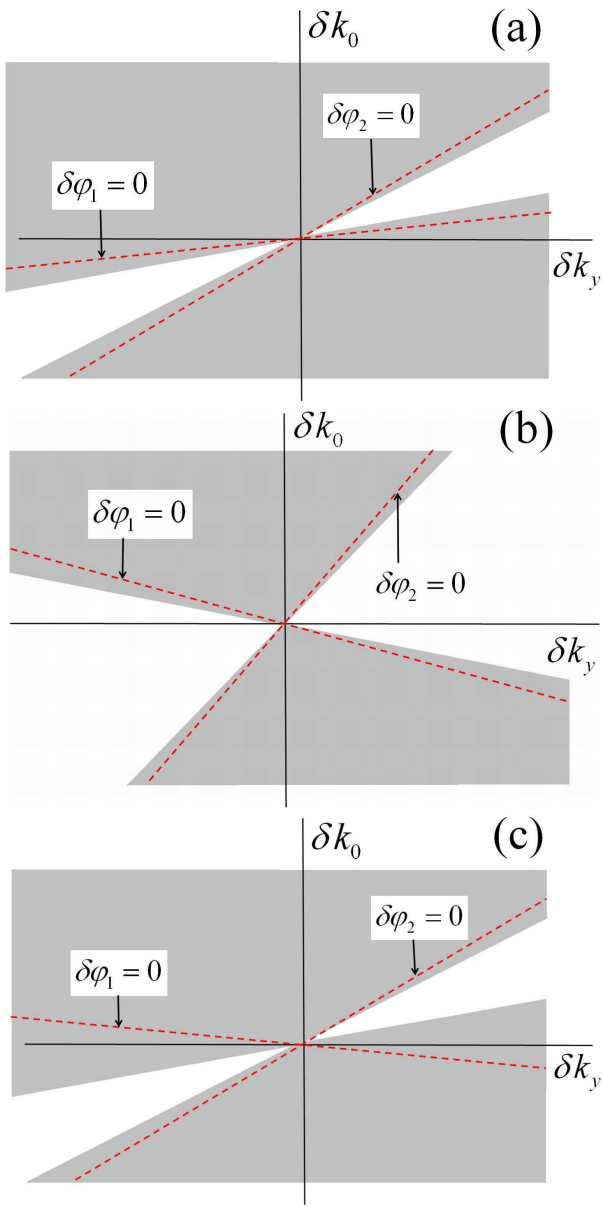}}
\caption{(color online) Transparency zones (gray) in the plane $\protect%
\omega (k_{x},k_{y})$ for: (a) a lattice composed of conventional
dielectrics only; (b) and (c) lattices containing both left- and
right-handed dielectrics. In (b), the touching point between the zones
provides a point-like transparent zone. In (c),
the contact of two zones is a point-like non-transparent gap. The phase
deviations $\protect\delta\protect\varphi_{1,2}$ vanish on the red dashed
lines.}
\label{Fig3}
\end{figure}

Equation~(\ref{eq38}) shows that the phase variations $\delta \varphi _{j}$
vanish on the lines 
\begin{equation}
\delta k_{0}=\frac{v_{jg}k_{yt}}{ck_{jt}}\delta k_{y}.  \label{eq39}
\end{equation}%
These lines lie in the transparency zones and intersect at the point where they touch. The line associated with the right-handed
dielectric lies in the first and the third quadrants, whereas the line
associated with the left-handed metamaterial lies in the second and the fourth
quadrants, as it is shown in Fig.~\ref{Fig3}. When the lattice is composed
of conventional dielectrics only, both lines $\delta \varphi _{1}(\delta
k_{y},\delta k_{0})=0$ and $\delta \varphi _{2}(\delta k_{y},\delta k_{0})=0$
lie in the first and the third quadrants, and the transparency zones touch
one another in a manner shown in Fig.~\ref{Fig3}(a). There are band gaps above
and below the frequency $\omega _{t}=ck_{0t}$, and these gaps are degenerate
in \textit{pointlike nontransparent zones (gaps)} at the frequency $\omega
=\omega_{t}$. Note that the zone structure of the lattice composed of both
left-handed dielectrics is similar to the one shown in Fig.~\ref{Fig3}(a), but
symmetrically reflected with respect to the $\delta k_0$ axis. 

The band structure can be significantly different when the lattice contains
both left- and right-handed dielectrics. In this case, there are
transparency zones above and below the frequency $\omega _{t}=ck_{0t}$, that
touch one another at the frequency $\omega =\omega _{t}$, forming a \textit{%
pointlike transparent zone} as it is shown in Fig.~\ref{Fig3}(b). This band
structure is similar to the energy spectrum of the charge carriers in
graphene, and manifests genuine Dirac points. It must be emphasized that for
such points to exist, the layers with both positive and negative refractive
indices must be present, but the average (over the period)
value of $n$ should not necessarily be equal to zero. Materials with
effective $\varepsilon$ and/or effective $\mu$ near zero have become
recently the subject of intensive investigation due to their unusual
transport properties, including the existence of conical singularities in
the band gap structures\cite{48,49,50,51,52}, however, this topic lies
outside the domain of our paper.

It is worth noting that the zone structures shown in Fig.~\ref{Fig3} have
been calculated for TE (\textit{p}-polarized) waves. In the case of \textit{s%
}-polarized fields, the dielectric permittivities $\varepsilon_i$ in Eq.~(%
\ref{eq36}) are replaced by the magnetic permeabilities $\mu_i$, therefore
the photonic band structure is different, while the coordinates of the
diabolic points are the same. This means that for a given frequency and
angle of incidence, the same sample could be transparent for \textit{s}%
-polarization and opaque \textit{p}-polarization, and vice versa. This
property can by utilized for the separation of polarizations.

\begin{figure}[tbh]
\centering \scalebox{0.4}{\includegraphics{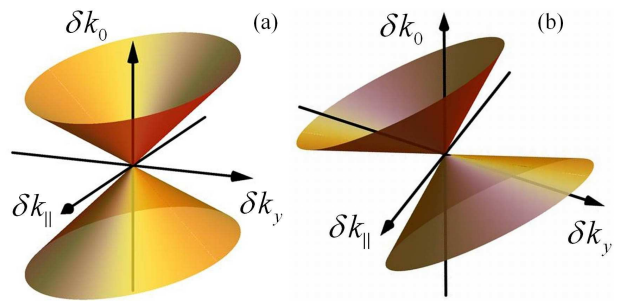}}
\caption{(color online) Surfaces $\protect\delta k_0(\protect\delta %
k_\parallel,\protect\delta k_y)$ for (a) the zone structure depicted in
Fig.~\ref{Fig3}(b) and (b) the zone structures shown in Figs.~\ref{Fig3}(a) and \ref{Fig3}(c).}
\label{Fig3a}
\end{figure}

The transparency zones depicted in Figs.~\ref{Fig3} are the projections onto
the plane $(k_{y},k_0)$ of the surface $\omega =\omega (k_{x},k_{y})$,
described by the dispersion equation (\ref{eq32a}) and shown in Fig.~\ref%
{Fig3a}. The zone structures, presented in Figs.~\ref{Fig3}(a), \ref{Fig3}(c), and \ref{Fig3}(b), correspond to the surfaces shown in Figs.~\ref{Fig3a}(a) and \ref{Fig3a}(b),
respectively. 

It is important to note that the presence of both left- and right-handed
dielectrics in the periodic structure is a \textit{necessary, but not
sufficient} condition for the graphene-like band structure to exist.
Generally speaking, the lines (\ref{eq39}) and the band edges do not
coincide, and can be widely separated. When the distance between the lines (%
\ref{eq39}) and the zone edges is large enough, as, for example, in Fig.~\ref%
{Fig3}(c), the zones shape is similar to that of a lattice formed by
right-handed dielectrics only [see Fig.~\ref{Fig3}(c)]. %
A detailed analysis of the zones shape is presented in
Appendix.

The fundamental qualitative difference between the zone shapes of mono- and
mix-lattices owes its origin to the strong dispersion, which is an inherent
characteristic of left-haded media. Ignoring this fact (\textquotedblleft
for simplicity\textquotedblright as this is sometimes done) leads to wrong
results for the zones structure in the vicinity of their %
contact.

Indeed, assuming a constant refractive index
$n_j(k_0)=\mathrm{const}$, the phase variation $\delta\varphi_j$ can be
written as 
\begin{equation}
\delta \varphi _{j}=\frac{d_{j}}{k_{jx}}\left( n_{j}^{2}k_{0}\delta
k_{0}-k_{y}\delta k_{y}\right) ,  \label{new1}
\end{equation}%
i.e., the phase variation vanishes on the line that lies in
the first and the third quadrants, independently of the sign of $n_j$.

\subsection{Graphene superlattice}

It follows from Eq.~(\ref{eqG1}) that in a graphene superlattice created by
an electrostatic potential with periodically-alternating values $u_1$ and $%
u_2$, the phase variations $\delta\varphi$ vanish on the lines 
\begin{equation}  \label{eqG2}
\delta w=\frac{\hbar^2 v_F^2k_{yt}}{w_t-u_j}\delta k_y,
\end{equation}
where $w_t$ and $k_{yt}$ are the coordinates of the zone %
touching point in the plane $(k_y,\,w)$. The slopes of
these lines have opposite signs when the energy $w_t$ lies between the
values $u_1$ and $u_2$ of the potential, i.e., when $\mathrm{sgn}%
[(w-u_1)(w-u_2)]=-1$. In this case, the zones touch each other either as it
is shown in Figs.~\ref{Fig3}(b) or \ref{Fig3}(c) (depending on the relation
between the layer thicknesses $d_1$ and $d_2$, and the %
touching point indices $p$ and $q$). In the opposite case,
when $\mathrm{sgn}[(w-u_1)(w-u_2)]=+1$, the zones structure is similar to
the one shown in Fig.~\ref{Fig3}(a), or symmetrically reflected with respect
to the ordinate axis. Thus, the point-like transparent zones (new Dirac
points) can appear in the graphene superlattice only in the energy range
between $u_1$ and $u_2$.

\section{Transmission near the Dirac points}

\begin{figure}[tbh]
\centering \scalebox{0.6}{\includegraphics{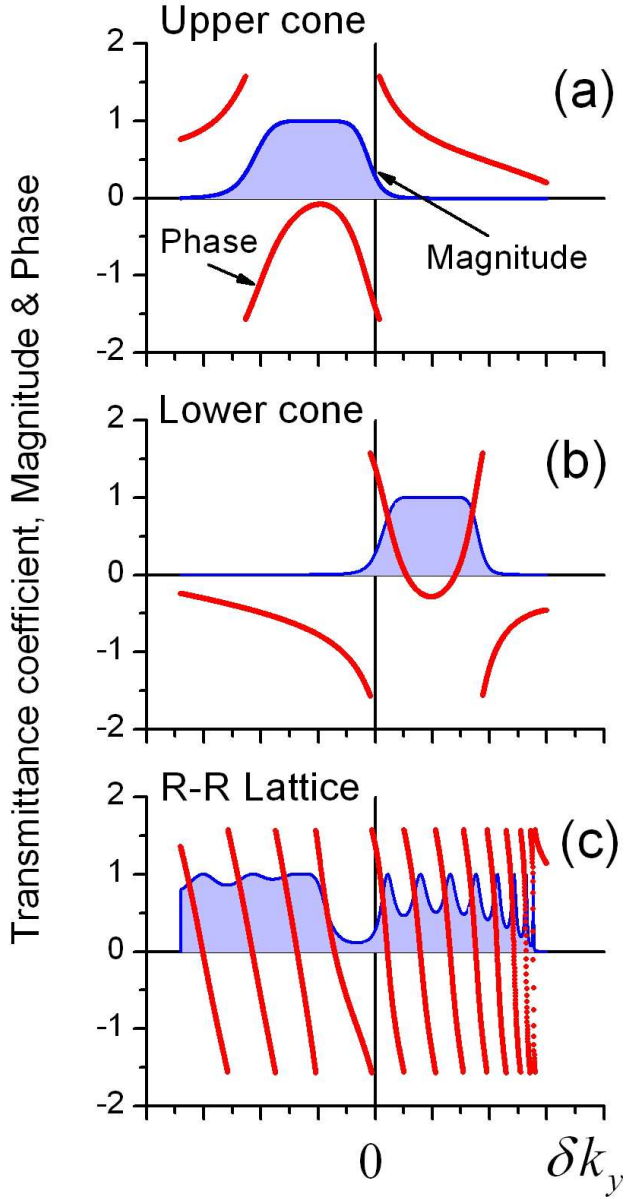}}
\caption{(color online) The magnitude (blue) and the phase (red) of the
complex transmission coefficient $t(k_{y})$ of a monochromatic wave
propagating through [(a) and (b)] finite periodic stacks of alternating left- and
right-handed dielectric layers and (c) a stack of normal dielectric layers. The
frequencies in (a) and (b) belong to the upper and lower cones in Fig.~%
\protect\ref{Fig2}, respectively. The $k_{y}$ component of the wavevector
determines the angle of incidence $\protect\theta=\arcsin{(ck_{y}/\protect%
\omega)}$. The parameters of the numerical simulations used here are: number
of periods $N=9$, $p=1$, $q=8$, $d_1=0.35D$, $\protect\varepsilon_1=n_1=1$, $%
|\protect\varepsilon_2|=0.8$, $|n_2|=2.5$, $|v_{2g}|=0.25c$, and $D|\protect%
\delta k_0|=0.05$.}
\label{Fig4}
\end{figure}

As it was mentioned in Introduction, the similarity between the energy
spectra of electromagnetic waves in homogeneous media, and Dirac
quasiparticles is only formal and physically meaningless, because the two
cones in the photon spectra are identical. To demonstrate that the singular
points (presented above) in the band gap structure of mixed periodic
dielectric samples possess the properties that make them %
\textit{bona fide} Dirac points, we considered the
transmission of a monochromatic wave of a frequency $\omega $ through a
finite stack of alternating left- and right-handed dielectric slabs. The
dependencies of the amplitudes and phases on the complex transmission
coefficients $t(k_{y})$ of the $y$-component of the wavevector [i.e., of the
angle of incidence $\theta=\arcsin (ck_{y}/\omega ) $] are shown in Figs.~%
\ref{Fig4}(a) and \ref{Fig4}(b), for two different frequencies belonging,
respectively, to the upper and the lower cones in Fig.~\ref{Fig2}. While the
amplitudes $|t(k_{y})|$ are similar, the phases $\beta =\arg t(k_{y})$
manifest quite distinct behaviors: 
\begin{equation}
\beta (k_{y})\simeq \pm b\left( k_{y}-k_{ym}\right) ^{2},  \label{eq40}
\end{equation}%
where opposite signs correspond to two different cones, and $k_{ym}$ is the
position of the phase $\beta(k_y)$ extremum. For comparison, the dependences
of $|t(k_{y})|$ and $\beta (k_{y})$ for a periodic stack of normal
dielectric layers are shown in Fig.~\ref{Fig4}(c). In this case, the phase is
a monotonic (close to linear) function of the angle of incidence at all
frequencies, and has no singularities at $k_{y}=k_{yt}$. 

\begin{figure}[tbh]
\centering \scalebox{0.6}{\includegraphics{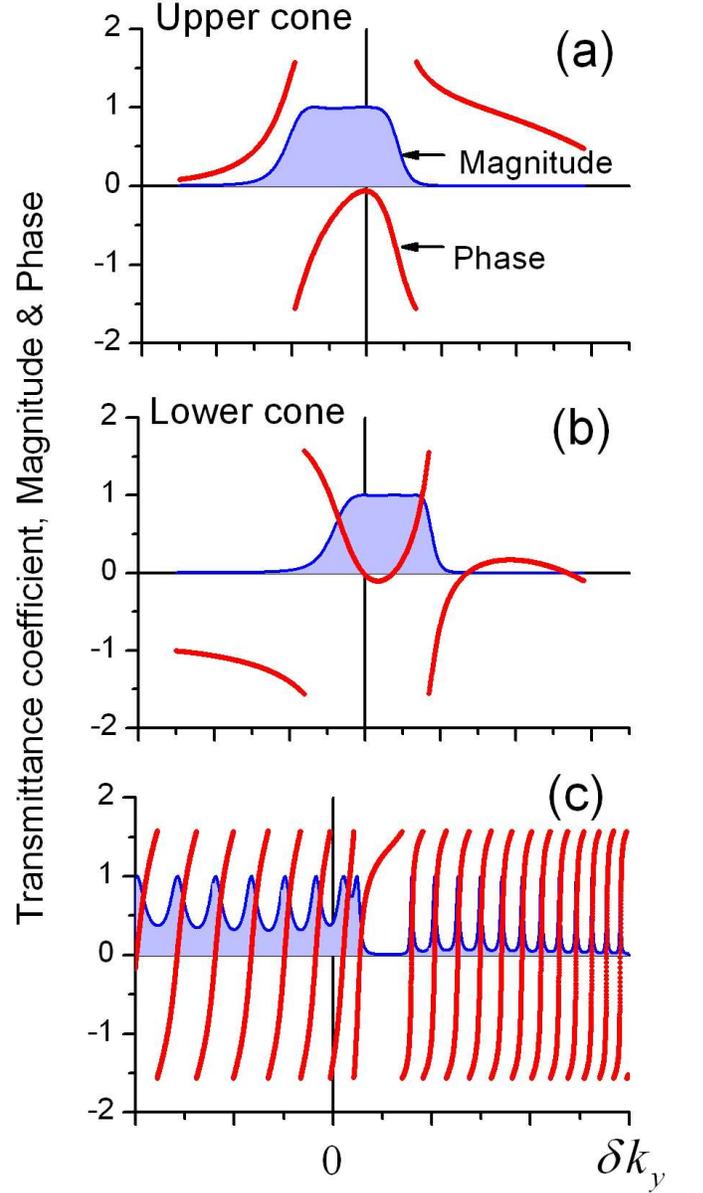}}
\caption{(color online) Magnitude (blue) and phase (red) of the complex
transmission coefficient $t(k_{y})$ of a monoenergetic quasiparticle
propagating through [(a) and (b)] a finite periodic set of alternating $p-n$ and $n-p$
junctions in graphene. The quasiparticle energies are (a) above  and (b) below
the energy of the zones touching $w_t$. The indexes
of the zones contact point are $p=1$, $q=5$, and the layer
thicknesses are $d_1=0.3D$ and $d_2=0.7D$. In the same lattice, the zones 
contact point, whose indexes are $p=3$ and $q=1$,
corresponds to a point-like gap, and the dependence $t(k_{y})$ looks like
the one for a periodic stack of normal dielectrics [see Fig.~\ref{Fig4}(c)].}
\label{Fig4a}
\end{figure}

Exactly the same parabolic dependence of the phase $\beta(k_y)$ of the
transmission coefficient is intrinsic to graphene superlattices, when the %
contact point corresponds to the point-like transparent
zone. The dependences of the amplitudes and phases of the complex
transmission coefficients $t(k_{y})$ on the $y$-component of the wavevector
are shown in Fig.~\ref{Fig4a} for two different energies: one above [see Fig.~%
\ref{Fig4a}(a)] and another below [see Fig.~\ref{Fig4a}(b)] the energy $w_t$ of the
zones contact. 

As it was mentioned above, the presence of both left- and right-handed
dielectrics, or alternating $p-n$ and $n-p$ junctions in graphene, is not
sufficient for a point of zone touching to be a point-like
transparency zone (Dirac point). In the same lattice, points of the zone-%
touching with different indexes $p$ and $q$ can be either
point-like gaps, as shown in Fig.~\ref{Fig4a}(c), or pointlike transparent
zones, i.e., Dirac points [see Figs.~\ref{Fig4a}(a) and \ref{Fig4a}(b)].

\begin{figure}[tbh]
\centering \scalebox{0.4}{\includegraphics{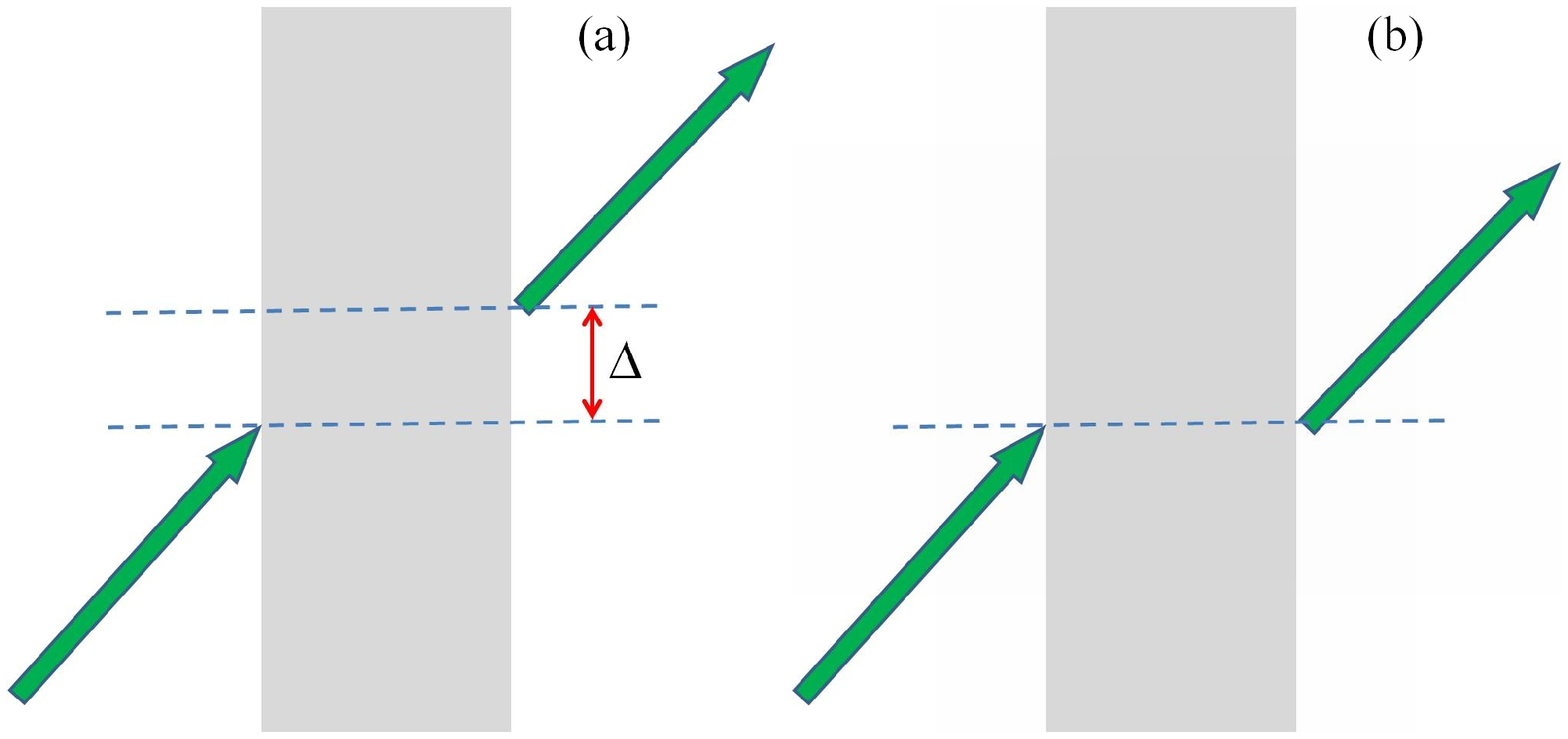}}
\caption{(color online) (a) A beam of light transmitted through a slab of a
normal dielectric is shifted along the surface. The shift of the beam
passing through a mixed periodic stack is much smaller, can be either
positive or negative, or even equal to zero, as shown in (b).}
\label{FigShift}
\end{figure}

A number of interesting effects emerge in the vicinity of the Dirac point of
the mixed periodic sample or graphene superlattice. Some of them are caused
by the parabolic dependence (\ref{eq40}) of the transmittance coefficient
phase on the angle of incidence (of $k_{y}$). Let us consider the
propagation of a monochromatic beam of light, bounded in the transverse
dimension (Gaussian beam, for instance), through the mixed stack of a
thickness $L$. Generally, a beam transmitted through a slab of a normal
dielectric is shifted along the surface, as it is schematically shown in
Fig.~\ref{FigShift}(a). 
The shift $\Delta $ is determined\cite{Artman} by evolution
of the phase $\beta $ of the transmission coefficient $t$: 
\begin{equation}
\Delta =-\left. \frac{d\beta (k_{y})}{dk_{y}}\right\vert _{k_{y}=k_{y0}},
\label{eq41}
\end{equation}%
where $k_{y0}$ is the tangential component of the wave vector of the central
ray in the incident beam. It is assumed in Eq.~(\ref{eq41}) that the angular
width $\Delta k_{y}$ of the incident beam is rather small. Since mixed
periodic stacks are characterized by the parabolic dependence $\beta (k_{y})$%
, Eq.~(\ref{eq40}), $d\beta (k_{y})/dk_{y}$ is small or even equal to zero, in
which case the longitudinal shift $\Delta $ is absent completely.

The absence of the longitudinal shift, when $\left.d\beta
(k_{y})/dk_{y}\right|_{k_y=k_{y0}}=0$, can also be observed in the so-called
near-zero-index metamaterials \cite{Kocaman} and in 1D periodic lattices
with birefringent materials \cite{Mandatori}. It is important to note that
in our case, this phenomenon has a different physical origin.

The parabolic dependence $\beta (k_{y})$ affects also the curvature of the
transmitted beam phase front, i.e., it shifts the focal plane of the
incident Gaussian beam. The value $\Delta _{f}$ of this shift is
proportional to the thickness $L$ of the mixed stack and is independent of
the stack position on the beam trajectory. The sign of the shift $\Delta
_{f} $ depends on the sign of the detuning of the beam frequency $\omega $
from the Dirac-point frequency $\omega _{t}$. The focal plane is shifted
forward (defocusing) when $\omega <\omega _{t}$, and backward (focusing)
when $\omega >\omega _{t}$.

\begin{figure}[tbh]
\centering \scalebox{0.3}{\includegraphics{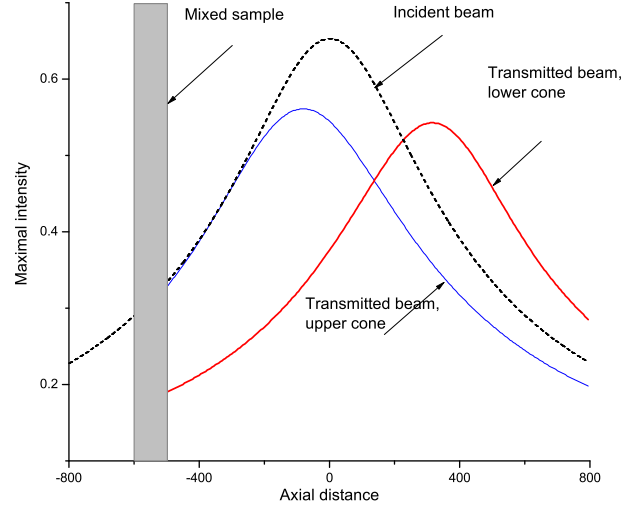}}
\caption{(color online) Focusing properties of a mixed periodic sample.
After the transmission through a mixed periodic sample (grey area), the
focus of the beam with a frequency belonging to the upper cone (blue curve)
is shifted to the left with respect to the beam propagating in free space
(black curve). The same sample shifts the focus of the beam with frequency
from the lower cone (red curve) in the opposite direction. The parameters of
the numerical simulations used here are the number of periods $N=100$, $p=2$, $%
q=5$, $d_1=0.4D$, $\protect\varepsilon_1=n_1=1$, $\protect\varepsilon_2=-0.5$%
, $n_2=-0.5$, $v_{2g}=-0.6c$, and $D|\protect\delta k_0|=0.3$.}
\label{Focus}
\end{figure}

This surprising focusing properties of mixed periodic samples are
demonstrated in Fig.~\ref{Focus}. Of the two beams with the frequencies
resting on two different dispersive cones, the one corresponding to the
upper cone is focused by the sample, blue curve, while the other (lower
cone) is defocused, red line. For comparison, the black curve presents the
intensity distribution in the beam propagated in free space. This phenomenon
is highly unusual by itself, and also reinforces the similarity of the %
contact points of the cones to a genuine ``optical'' Dirac
point: different cones are not identical, and represent objects with
distinct physical properties --- a sort of optical particle-antiparticle
pair. 

When the frequency $\omega$ of the incident beam coincides
with a point-like gap of the corresponding infinite structure, i.e., $%
\omega=\omega_t$, one would expect an exponential decay of the transmitted
beam intensity $I_{\mathrm{tr}}$ as a function of the mixed stack thickness $%
L$: $I_{\mathrm{tr}}(L)\propto\exp(-\gamma L)$. However, the well-pronounced
constant asymptotic of the function $L\cdot I_{\mathrm{tr}}(L)$ (Fig.~\ref%
{Intensity}, red line), demonstrates an anomalously high intensity, decaying
as $1/L$. Such a diffusion-like dependence is one of the consequences of the
linear, Dirac cone-like dispersion. In two-dimensional photonic crystals
with triangular lattices of normal dielectric rods, this phenomenon was
predicted in Ref.~\onlinecite{9}. In layered structures, it could take place
only in the presence of left-handed elements. 
\begin{figure}[tbh]
\centering \scalebox{0.22}{\includegraphics{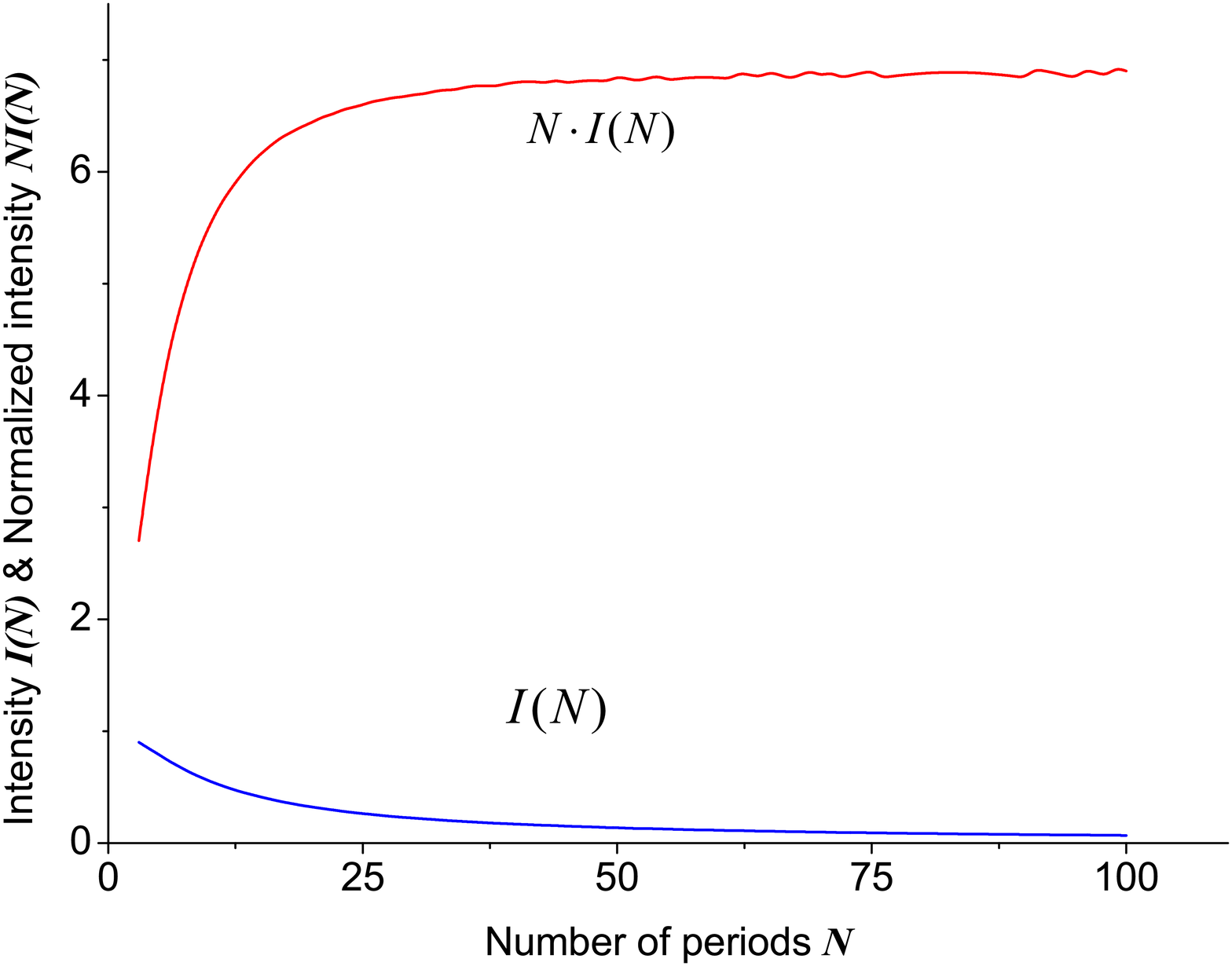}}
\caption{(color online) Intensity $I(N)$ (blue line) of the transmitted beam
as a function of the distance of propagation inside a mixed stack. The
distance is measured in the numbers of periods $N$. The frequency $\protect%
\omega$ of the incident beam belongs to a point-like transparency zone
(Dirac point) of the corresponding infinite structure, $\protect\omega=%
\protect\omega_{t}$. The large-$N$ asymptotic of the red line, $NI(N)$, is
constant, which means that the intensity is inversely proportional to the
distance (diffusion-like dependence). The parameters of the numerical
simulations here are: $p=1$, $q=8$, $d_1=0.35D$, $\protect\varepsilon_1=n_1=1
$, $\protect\varepsilon_2=-0.8$, $n_2=-2.5$, and $v_{2g}=-0.25c$.}
\label{Intensity}
\end{figure}

\begin{figure}[tbh]
\centering \scalebox{0.4}{\includegraphics{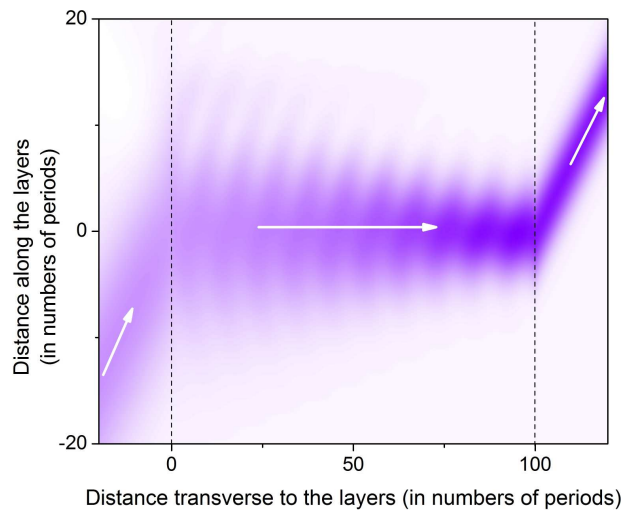}}
\caption{(color online) Spatial distribution of the energy flux of the beam
propagating (from left to right) through the mixed sample. 
The angle of incidence $\simeq 34^\circ$. The boundaries of
the sample are marked by the vertical dashed lines. The ``center of
gravity'' of the flux oscillates in the transverse (along the $y$-axis)
direction. The parameters of the numerical simulations used here are the number
of periods $N=100$, $p=4$, $q=10$, $d_1=0.4D$, $\protect\varepsilon_1=n_1=1$%
, $\protect\varepsilon_2=-0.5$, $n_2=-1.5$, $v_{2g}=-0.2c$, $D\protect\delta %
k_0=-0.07$, and $D\protect\delta k_y=0.5$.}
\label{Zitter1}
\end{figure}

\begin{figure}[tbh]
\centering \scalebox{0.33}{\includegraphics{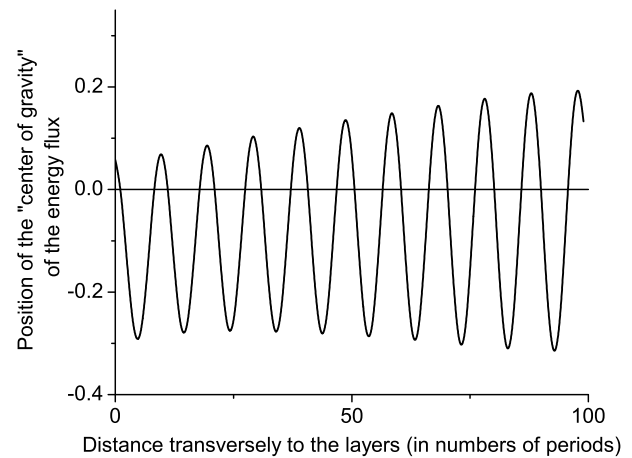}}
\caption{Oscillations of the transverse coordinate of the ``center of
gravity'' of the beam propagating in the mixed sample.}
\label{Zitter2}
\end{figure}

Because of the similarity between the energy spectrum of relativistic
electrons and the frequency band structure of a mixed periodic dielectric
structure, it is natural to assume that the light beam propagation into
mixed samples can be accompanied by a \textit{Zitterbewegung}-like
phenomenon. Indeed, the spatial distribution of the energy flux inside the
mixed sample manifests a \textit{trembling motion}: the ``center of
gravity'' of the flux oscillates in the transverse direction (along the $y$%
-axis). In Fig.~\ref{Zitter1}, the spatial distribution of the energy flux
inside the mixed sample is shown. The oscillatory motion of the center of
gravity $I_{c}(y)$ is clearly seen in Fig.~\ref{Zitter2}. Note that Fig.~\ref%
{Zitter1} is similar to Fig. 3 in Ref.~\onlinecite{Thaler}, where the
probability function of a moving electron (solution of the Dirac equation)
is shown. 

Figure~\ref{Zitter1} also demonstrates the two above-mentioned effects: the
absence of the longitudinal shift (the energy propagates normally to the
sample boundary, whereas the angle of incidence of the beam
is far from the normal) and the focusing of the beam.

\begin{figure}[tbh]
\centering \scalebox{0.4}{\includegraphics{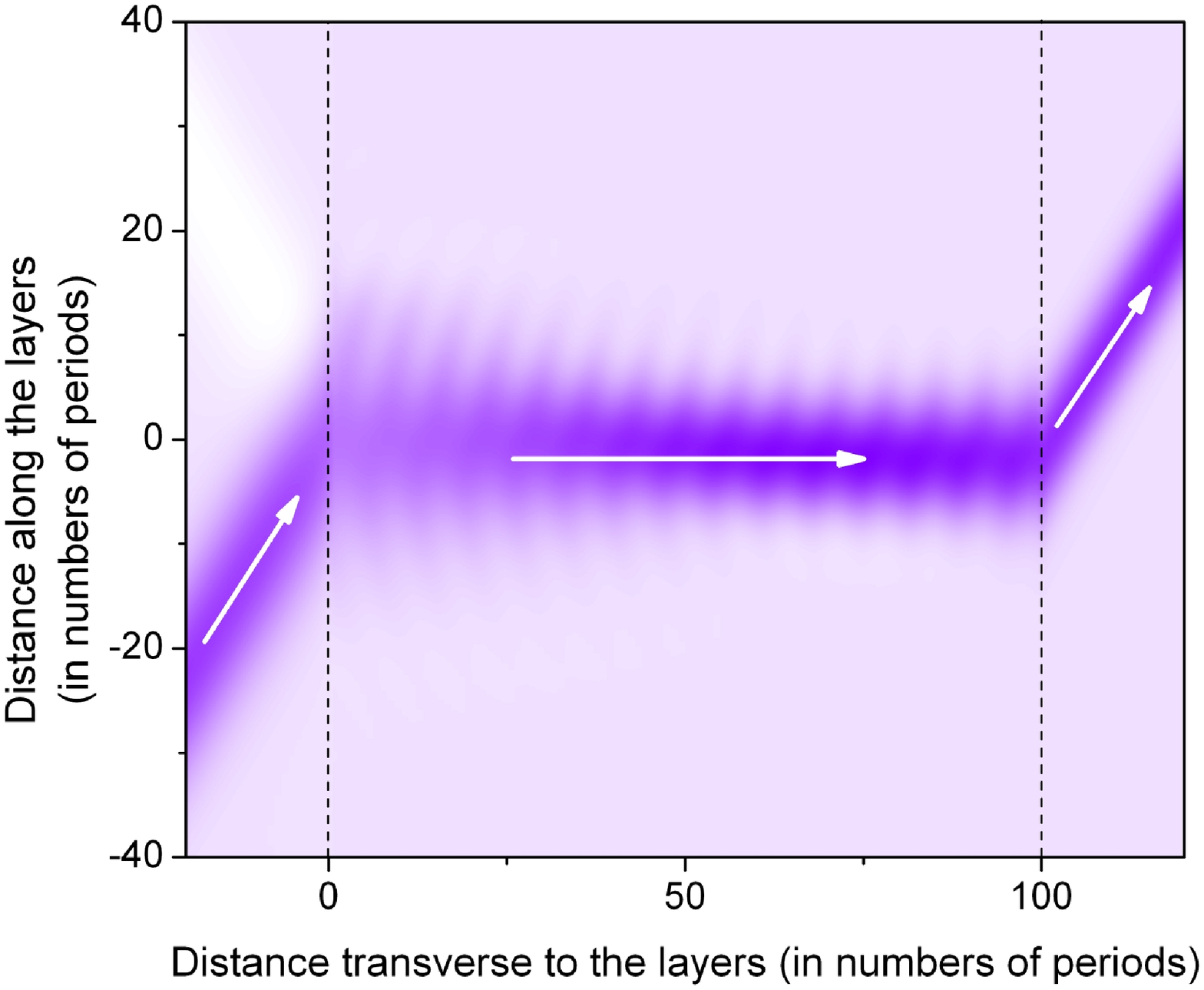}}
\caption{(color online) Transmission of the Gaussian beam of charge carriers
through a periodic set of $p-n$ and $n-p$ junctions in graphene near a
point-like gap. The angle of incidence $\simeq 48^\circ$.
The sample boundaries are marked by the vertical dashed lines. The
parameters of the numerical simulations used here are the number of periods $%
N=100$, $p=1$, $q=5$, $d_1=0.3D$, $u_1=-u_2=20\hbar v_F/D$, $\protect\delta %
w=-0.3\hbar v_F/D$, and $D\protect\delta k_y=0.2$.}
\label{GPulse}
\end{figure}

All these effects --- the absence of the longitudinal shift, the focusing of
the beam, and the Zitterbewegung phenomenon --- are also seen in graphene
superlattices. As an example, the current density distribution in the
Gaussian beam of monoenergetic charge carriers propagating through a finite
graphene superlattice is depicted in Fig.~\ref{GPulse}. 

Let us now compare the wave propagation through a mixed dielectric slab or a
graphene superlattice with the transmission of an electromagnetic wave
through a plate made of a \textit{homogeneous} (right- or left-handed)
dielectric. In a small vicinity of the normal angle of incidence, the
dependence $\beta(k_{y})$ of the transmission coefficient phase $\beta $ on
the tangential component of the wave number $k_{y}$ (the dependence on the
angle of incidence) has the same parabolic form as discussed above [see Eq.~(%
\ref{eq40})], with the only difference that at homogeneous dielectric
plates, $k_{yt}=0$ by definition: 
\begin{equation}  \label{eq42}
\beta(k_y)=\pm bk_y^2.
\end{equation}
Here the plus sign corresponds to right-handed dielectrics, while the minus
sign corresponds to left-handed dielectrics. In the first case, the
refraction is positive, i.e., the incident beam and the beam inside the
plate lie on opposite sides from the normal; whereas in the second case, the
beams lie on the same side of the normal (negative refraction). Therefore,
the shift $\Delta $ is positive for plates of a right-handed dielectric, and
negative for left-handed dielectrics. In other words, in normal samples, $%
k_{y}$ and $\Delta $ have the same signs, while in metamaterials the signs
are opposite. Amazingly, these two situations, each inherent to \textit{%
different} kinds of homogeneous materials, can be observed in the \textit{%
same} mixed periodically-layered sample, and graphene superlattice. Indeed,
depending on which side of the Dirac point the frequency lies, the
refraction is either positive (upper cone), or negative (lower cone). In
this regard, it is more appropriate to refer to these two cones as a
``medium -- anti-medium'' pair, rather than ``particle--antiparticle'', as
in homogeneous graphene.

\section{Conclusions}

We have shown that some of the exotic properties of charge transport in
graphene can be reproduced in the propagation of light through layered
dielectric samples. Similarities and distinctions between Maxwell and Dirac
equations, and between the corresponding boundary conditions have been
studied. Although the equations for the real electric and magnetic fields
are essentially different from those for the Dirac electrons, under some
conditions they can be reduced to similar form. For example, Eq.~(\ref{eq4})
for the complex combinations given by Eq.~(\ref{eq3}) coincides with the
Dirac equations (\ref{eq2}). Therewith, the role of the refractive index
in graphene is played by the difference between properly normalized values
of the Fermi energy and the external electrostatic potential [Eq.~(\ref{eq5}%
)]. The boundary conditions for a Dirac quasiparticle incident on a plane
separating two areas with different potentials, and for an electromagnetic
wave propagating through an interface between two layers of homogeneous
dielectrics are, generally speaking, different. They coincide only when the
impedances of the layers are equal, and the direction of the propagation is
normal to the boundary. It means that at normal incidence, any junction in
graphene is analogous to a contact between two perfectly matched dielectrics
and therefore is absolutely transparent to normally-incident Dirac
electrons. This provides a more intuitive insight into the physics of the
Klein paradox in graphene.

The analytical and numerical analysis of the photonic band gap structures of
infinite periodically-layered systems reveals an infinite number of the
so-called ``diabolic points'' (singular points of contact of two
transparency zones) in the $(\omega,\theta)$ spectral diagrams (examples are
shown in Fig.~\ref{Fig3}). A distinction needs to be drawn between two types
of these singularities: point-like transparency zones, like in Figs.~\ref%
{Fig3}(a) and \ref{Fig3}(c), and pointlike gaps in the spectrum, similar to the
one presented in Fig.~\ref{Fig3}(c). Although all three pictures in Fig.~\ref%
{Fig3} are topologically equivalent, the transport properties of the
corresponding \textit{finite} periodic stacks of layers differ drastically
in the vicinities of these points. Waves with frequencies lying on opposite
sides of the singularities of the first type propagate through the samples
in similar ways. In the same time, when two touching spectral cones form a point-like gap, the electromagnetic radiation
interacts with the same sample differently, depending to which cone its
frequency belongs. Studies of the propagation of beams of light show that
only the diabolic points of this type posses the properties of genuine Dirac
points. We demonstrate that in mono-type layered structures (i.e., in those
built of either normal or left-handed dielectrics) just the diabolic point
of the first type can exist, and conical, Dirac-type singularities appear
only in mixed (with alternating left- and right-handed layers) samples. This
is in contrast to two-dimensional media where Dirac points were discovered
in various types of photonic crystals with normal dielectric elements. It is
important to note that the physical nature of the Dirac points that we
consider is different from that in systems with zero average value of the
refractive index: in mixed layered structures, they are due to the specific
strong dispersion (phase and group velocities have different signs) inherent
in the elements with negative refraction.

Although the angular dependencies of the transmission and reflection
coefficients from a single interface in layered dielectrics and graphene
superlatices are different [compare formulas Eqs.~(\ref{eq18}) and (\ref%
{eq24})], the spectral properties of these two structures are conceptually
identical and entail similar features in the light and charge transport.
Considering, as examples, the transmission of the Gaussian monochromatic
beams of light and monoenergetic Dirac electrons through the corresponding
(dielectric or graphene) samples we predict the following
Dirac-point-induced effects: (i) two touching Dirac cones
influence the propagation of a beam \ in different ways: the beam is focused
when the frequency (energy) belongs to the upper cone, and is defocused at
frequencies (energies) lying in the lower one, (ii) the transverse shift of
the beam is anomalously small or even zero, (iii) the decay of the intensity
at forbidden frequencies is diffusion-like, and (iv) a spatial analog of the
Zitterbewegung effect (i.e., trembling motion of the ``center of gravity''
of the energy flux) is observed in periodically layered dielectric
structures with nonzero average refractive indices and in graphene
superlatices.

\section{Acknowledgments}

This work was supported in part by JSPS-RFBR Grant No. 12-02-92100, RFBR
Grant No. 11-02-00708, ARO, RIKEN iTHES Project, MURI Center for Dynamic Magneto-Optics, Grant-in-Aid for Scientific Research (S), MEXT
Kakenhi on Quantum Cybernetics, the JSPS via its FIRST program,  and
by the Israeli Science Foundation (Grant No. 894/10).

\appendix*

\section{Boundaries of the transparency zones of photonic structures}

Boundaries of the transparency zones in the plane $(\delta k_y, \delta k_0)$
in the vicinity of the zones touching point are defined by the equation 
\begin{equation}  \label{A1}
\delta\varphi_1^2+\delta\varphi_2^2+2a\delta\varphi_1\delta\varphi_2=0,
\end{equation}
where $\delta\varphi_j$ are defined by Eqs.~(\ref{eq38}). It follows from
Eq.~(\ref{A1}), that 
\begin{equation}  \label{A2}
\delta\varphi_2=\left(-a\pm\sqrt{a^2-1}\right),
\end{equation}
where 
\[
a=\frac{1}{2}\left(\frac{k_{1x}\varepsilon_2}{k_{2x}\varepsilon_1}+\frac{%
k_{2x}\varepsilon_1}{k_{1x}\varepsilon_2}\right). 
\]
This equation, using Eq.~(\ref{eq38}), can be presented in the form: 
\begin{equation}  \label{A3}
\delta k_0^{(\pm)}=\delta k_y\frac{k_{yt}}{k_{0t}}\frac{d_2k_{1x}-d_1k_{2x}%
\left(-a\pm\sqrt{a^2-1}\right)}{d_2k_{1x}\frac{c}{v_{g2}}-d_1k_{2x}\frac{c}{%
v_{g1}}\left(-a\pm\sqrt{a^2-1}\right)}.
\end{equation}

When the photonic structure is formed by either right- or left-handed layers
only, i.e., $a>0$ and $\mathrm{sgn}(v_{g1}/v_{g2})=1$, both lines $\delta
k_{0}^{(\pm )}(\delta k_{y})$ lie in the same quadrants and the zones
touching point forms a point-like gap, as it is shown, for instance, in Fig.~%
\ref{Fig3}a. In the mixed structure which contains both left- and
right-handed layers, i.e., $a<0$ and $\mathrm{sgn}(v_{g1}/v_{g2})=-1$, the
situation is quite different. The denominator in Eq.~(\ref{A3}) has the same
sign for both $\delta k_{0}^{(+)}$ and $\delta k_{0}^{(-)}$, while the signs
of the numerator \textit{can} be different. Whether the sign of $%
(\delta k_{0}^{(+)}/\delta k_{0}^{(-)})$ is equal to $-1$ or $+1$ depends on
the values of the parameters. The zones structures in the first 
[$\mathrm{sgn}\left(\delta k_{0}^{(+)}/\delta k_{0}^{(-)}\right)=-1$]  and the second [$\mathrm{sgn}\left(\delta k_{0}^{(+)}/\delta k_{0}^{(-)}\right)=+1$] cases are similar to ones
shown in Fig.~\ref{Fig3}b and Fig.~\ref{Fig3}c, respectively.

Thus, the zones touching point presents the point-like transparent zone when
the following inequalities hold: 
\begin{eqnarray}  \label{A4}
\frac{d_2k_{1x}}{d_1k_{2x}}<|a|+\sqrt{a^2-1},  \nonumber \\
\frac{d_2k_{1x}}{d_1k_{2x}}>|a|-\sqrt{a^2-1}.
\end{eqnarray}

Using the definition of the parameter $a,$ and Eq.~(\ref{eq34}) these
inequalities can be written as 
\begin{equation}
\left( \frac{d_{1}}{d_{2}}\right) ^{3}\left( \frac{q}{p}\right) ^{2}\left( 
\frac{\varepsilon _{1}}{|\varepsilon _{2}|}\right) <1<\left( \frac{d_{1}}{%
d_{2}}\right) \left( \frac{|\varepsilon _{2}|}{\varepsilon _{1}}\right) ,
\label{A5}
\end{equation}%
when 
\[
\left( \frac{d_{2}}{d_{1}}\right) \left( \frac{p}{q}\right) \left( \frac{%
|\varepsilon _{2}|}{\varepsilon _{1}}\right) >1,
\]%
and reverse to Eq.~(\ref{A5}) inequalities, when 
\[
\left( \frac{d_{2}}{d_{1}}\right) \left( \frac{p}{q}\right) \left( \frac{%
|\varepsilon _{2}|}{\varepsilon _{1}}\right) <1.
\]%
As it follows from Eq.~(\ref{eq34}), the range of allowed values
of the parameters is restricted by the condition 
\begin{equation}
\left( \frac{d_{2}}{d_{1}}\right) \left( \frac{p}{q}\right) \left( \frac{%
|n_{2}|}{n_{1}}\right) >1,  \label{A6}
\end{equation}%
when $n_{1}>|n_{2}|$, and by the reverse inequality when $n_{1}<|n_{2}|$.

All these inequalities select a region in the
4-dimensional space of parameters $d_{1}/d_{2}$, $p/q$, $\varepsilon
_{1}/|\varepsilon _{2}|$, and $n_{1}/|n_{2}|$, where the vicinity of the
zones contact point is characterized by the Dirac-like spectrum.

\begin{figure}[tbh]
\centering \scalebox{0.35}{\includegraphics{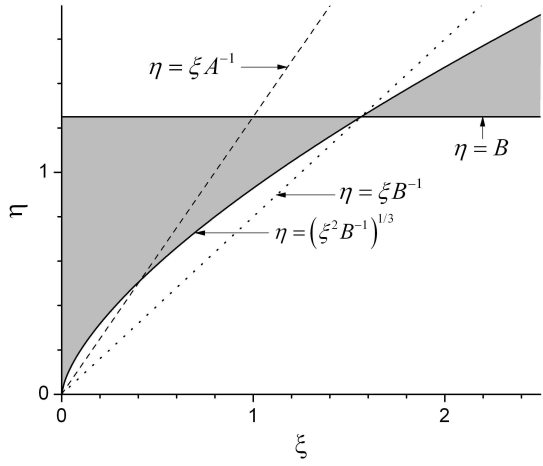}}
\caption{Plot of the photonic structure parameters $\protect\xi=d_1/d_2$ and 
$\protect\eta=p/q$. The vicinity of the zones contact point is characterized by a
Dirac-like spectrum when the structure parameters belong in the gray region
below the dashed line (if $A=n_1/|n_2|>1$) or above this line (if $%
A=n_1/|n_2|<1$).}
\label{FigA}
\end{figure}

Any metamaterial-dielectric pair is characterized by its own values of the
parameters $\varepsilon _{1}/|\varepsilon _{2}|$ and $n_{1}/|n_{2}|$.
Therefore, considering these parameters as given, one can define\ the
corresponding area in the two-dimensional space of parameters $\xi =p/q$ and 
$\eta =d_{1}/d_{2}$ (remainder: $p$ and $q$ are integer numbers).
Introducing the notations $A=n_{1}/|n_{2}|$ and $B=\varepsilon
_{1}/|\varepsilon _{2}|$ we can write the inequalities Eqs.~(\ref{A5}) and (%
\ref{A6}) as 
\begin{eqnarray}
B\eta ^{3} &<&\xi ^{2},\hspace{3mm}\eta <B,\hspace{3mm}B\eta <\xi , 
\nonumber  \label{A7} \\
B\eta ^{3} &>&\xi ^{2},\hspace{3mm}\eta >B,\hspace{3mm}B\eta >\xi ,
\end{eqnarray}%
and 
\begin{eqnarray}
\xi  &>&A\eta ,\hspace{3mm}\mathrm{if}\hspace{1mm}A>1,  \nonumber  \label{A8}
\\
\xi  &<&A\eta ,\hspace{3mm}\mathrm{if}\hspace{1mm}A<1.
\end{eqnarray}%
Equation (\ref{A8}) means that the line $\eta =A^{-1}\xi $ divides the plane $%
(\xi ,\eta )$ in two parts. The allowed values of variables $\xi $
and $\eta $ lie below this line if $A>1$, and above the line if $A<1$.
Inequalities Eq.~(\ref{A7}) bound an area between the line $\eta =B$ and
the curve $\eta =\left( x^{2}/B\right) ^{1/3}$. The region where a touching
point of the zones presents the genuine Dirac point is defined by the
intersection of these two areas, as it is shown in Fig.~\ref{FigA}.

Note, that the group velocities $v_{g1}$ and $|v_{g2}|$ are not involved in
this analysis. These velocities define the angle of opening of the Dirac
spectrum, but not the fact of its existence.

Usually, metamaterials exhibit their left-handed properties in a rather
narrow frequency range. Because the above-mentioned region is defined only
by  the relation between the layer thicknesses $d_{1}$ and $d_{2}$, one can
tune the Dirac point frequency $\omega _{t}=ck_{0t}$ by an appropriate
choice of the structure period $D$.

\begin{thebibliography}{99}
\bibitem{1} V.G. Veselago, Sov. Phys. Usp. \textbf{10}, 509 (1968).

\bibitem{2} P.R. Wallace, Phys. Rev. B \textbf{71}, 622 (1947).

\bibitem{3} I. Bialynicki-Birula, Progress in Optics, XXXVI (1996).

\bibitem{4} I. Bialynicki-Birula, Z. Bialynicka-Birula, J. Phys. A: Math. Theor. \textbf{46}, 053001 (2013); arXiv: 1211.2655
(2012).

\bibitem{5} F.D.M. Haldane, S. Raghu, Phys. Rev. Lett. \textbf{100}, 013904
(2008).

\bibitem{6} L.-G. Wang, Z.-G. Wang, J.-X. Zhang, S.-Y. Zhu, Opt. Lett. 
\textbf{34}, 1510 (2009).

\bibitem{7} T. Ochiai, M. Onoda, Phys. Rev. B. \textbf{80}, 155103 (2009).

\bibitem{7a} K.Y. Bliokh, A.Y. Bekshaev, and F. Nori, New J. Phys. \textbf{15%
}, 033026 (2013).

\bibitem{Shytov} A.V. Shytov, M.S. Rudner, and L.S. Levitov, Phys. Rev.
Lett. \textbf{101}, 156804 (2008).

\bibitem{Garcia} J.L. Garcia-Pomar, A. Cortijo, M. Nieto-Vesperinas, Phys.
Rev. Lett. \textbf{100}, 236801 (2008).

\bibitem{Darancet} P. Darancet, V. Olevano, and D. Mayou, Phys. Rev. Lett. 
\textbf{102}, 136803 (2009).

\bibitem{Rozhkov} A.V. Rozhkov, G. Giavaras, Y.P. Bliokh, V. Freilikher, F.
Nori, Phys. Rep. \textbf{503}, 77 (2011).

\bibitem{Chen} X. Chen, P.-L.Zhao, X.-J. Lu, arXiv:1111.1753 (2011); %
X. Chen, X.-J. Lu, Y. Ban, and C.-F. Li, J. Optics
\textbf{15}, 033001 (2013).

\bibitem{8} Y.P. Bliokh, V. Freilikher, S. Savel'ev, F. Nori, Phys. Rev. B 
\textbf{79}, 075123 (2009).

\bibitem{A} R. Polles, A. Moreau, G. Granet, Opt. Let.
\textbf{35}, 3237 (2010).

\bibitem{B}  H. Benisty, N. Piskunov, P.  Kashkarov, O.
Khayam, Phys. Rev. A \textbf{84}, 063825 (2011).

\bibitem{C} H. Benisty, O. Khayam, IEEE J. Quant.
Electronics \textbf{47}, 204 (2011).

\bibitem{D} H. Benisty, Phys. Rev. B \textbf{79}, 155409
(2009).

\bibitem{9} R. Sepkhanov, Ya. Bazaliy, C. W. J. Beenakker, Phys. Rev. A 
\textbf{75}, 063813 (2007).

\bibitem{10} M. Plihal, A. Shambrook, A. A. Maradudin, Opt. Commun. \textbf{%
80}, 199 (1991).

\bibitem{11} M. Plihal, A.A. Maradudin, Phys. Rev. B \textbf{44}, 8565
(1991).

\bibitem{12} D. Cassagne, C. Jouanin, D. Bertho, Phys. Rev. B \textbf{52},
R2217 (1995); \textbf{53}, 7134 (1996); Appl. Phys. Lett. \textbf{70}, 289
(1997).

\bibitem{15} F. Gadot, A. Chelnokov, A.D. Lustrac, P. Crozat, J.-M.
Lourtioz, D. Cassagne, C. Jouanin, Appl. Phys. Lett. \textbf{71}, 1780
(1997).

\bibitem{16} L. Martinez, A. Garcia-Martin, P. Postigo, Opt. Express \textbf{%
12}, 5684 (2004).

\bibitem{17} K. Busch, G. von Freymann, S. Linden, S. Mingaleev, L.
Tkeshelashvili, M. Wegener, Phys. Reports \textbf{444}, 101 (2007).

\bibitem{18} F.D.M. Haldane, S. Raghu, Phys. Rev. A \textbf{78}, 033834
(2008).

\bibitem{19} M. Diem, T. Koschny, C.M. Soukoulis, Physica B \textbf{405},
2990 (2010).

\bibitem{20} X. Zhang, Phys. Rev. Let \textbf{100}, 113903 (2008).

\bibitem{21} Q. Liang, Y. Yan, J. Dong, Opt. Lett. \textbf{36}, 2513 (2011) .

\bibitem{22} F. Dreisow, M. Heinrich, R. Keil, A. T\"{u}nnermann, S. Nolte,
S. Longhi, A. Szameit, Phys. Rev. Let. \textbf{105}, 143902 (2010).

\bibitem{23a} T. Ando, T. Nakanishi, and R. Saito, J. Phys.
Soc. Japan \textbf{67}, 2857 (1998).

\bibitem{23b} K.Y. Bliokh, Phys. Lett. A, \textbf{344}, 127
(2005).

\bibitem{23} R. Sepkhanov, A. Ossipov, C. W. J. Beenakker, Eur. Phys. Lett. 
\textbf{85}, 14005 (2009).

\bibitem{24} O. Peleg, G. Bartal, B. Freedman, O. Manela, M. Segev, D. N.
Christodoulides, Phys. Rev. Lett. \textbf{98}, 103901 (2007).

\bibitem{25} M. Rechtsman, Y. Plotnik, D. Song, M. Heinrich, J. M. Zeuner,
S. Nolte, N. Malkova, J. Xu, A. Szameit, Z. Chen, M. Segev, arXiv:1210.5361
(2012).

\bibitem{25a} M.C. Rechtsman, Y. Plotnik, J.M. Zeuner, A. Szameit, M. Segev,
arXiv:1211.5683 (2012).

\bibitem{26} X. Huang, Y. Lai, Z. H.Hang, H. Zheng, C. T. Chan, Nat. Mater. 
\textbf{10}, 582 (2011).

\bibitem{27} K. Sakoda and  H-F. Zhou, Opt. Express \textbf{18}, 27371 (2010); 
\textbf{19}, 13899 (2011); K. Sakoda, Opt. Express \textbf{20}, 3898 (2012); 
\textbf{20}, 9925 (2012); \textbf{20} 25181, (2012); J. Opt. Soc. Am. B 
\textbf{29}, 2770 (2012).

\bibitem{34} S. Matsuzawa, K. Sato, Y. Inoue, T. Nomura, IEICE Trans.
Electron. \textbf{E89-C}, 1337 (2006).

\bibitem{36} I. Nefedov, S. Tretyakov, Phys. Rev. E \textbf{66}, 036611
(2002).

\bibitem{37} L. Wu, S. He, L. Shen, Phys. Rev. B \textbf{67}, 235103 (2003).

\bibitem{38} D. Bria, B. Djafari-Rouhani, A. Akjouj, L. Dobrzynski, J.
Vigneron, E. El Boudouti, A. Nougaoui, Phys. Rev. E \textbf{69}, 066613
(2004).

\bibitem{Japan}  M. Tashima and N. Hatano, arXiv:1306.0297 (2013).

\bibitem{39} M. Katsnelson, K. S. Novoselov, A. Geim, Nat. Phys. \textbf{2},
620 (2006).

\bibitem{41} V. Cheianov, V. Fal'ko, B. Altshuler, Science \textbf{315},
1252 (2007).

\bibitem{42} J. B. Pendry, Phys. Rev. Lett. \textbf{85}, 3966 (2000).

\bibitem{43} A. Young, P. Kim, Nat. Phys. \textbf{5}, 222 (2009).

\bibitem{Bliokh} Y.P. Bliokh, V. Freilikher, F. Nori, Phys. Rev. B \textbf{81%
}, 075410 (2010).

\bibitem{Allian} P.E. Allain and J.N. Fuchs, Eur. Phys. J. B \textbf{83},
301 (2011).

\bibitem{45} P. Yeh, A. Yariv, .C.-S. Hong, J. Opt. Soc. Am. \textbf{97},
423 (1977).

\bibitem{Berry} M.V. Berry and M.R. Jeffrey, Progress in Optics \textbf{50},
13 (2007).

\bibitem{48} J. Li, L. Zhou, C.T. Chan, P. Sheng, Phys. Rev. Lett. \textbf{90%
}, 083901 (2003).


\bibitem{49} X. Huang, Y. Lai, Z. H. Hang, H. Zheng, C. T. Chan, Nature Mat. 
\textbf{10}, 582 (2011).

\bibitem{Artman} K. Artmann, Ann. Phys. Lpz. \textbf{2}, 87
(1948).

\bibitem{50} M.G. Silveirinha, N. Engheta, Phys. Rev. Lett. \textbf{97},
157403 (2006).

\bibitem{51} N.C. Panoiu, R.M. Osgood, Jr., S. Zhang, and S.R.J. Brueck, J. Opt.
Soc. Am. B \textbf{23}, 507 (2006).

\bibitem{52} S. Feng, Phys. Rev. Lett. \textbf{108}, 193904 (2012).

\bibitem{Kocaman} S. Kocaman, M.S. Aras, P. Hsieh, J.F. McMillan, C.G.
Biris, N.C. Panoiu, M.B. Yu, D.L. Kwong, A Stein, and W. Wong, Nat.
Photonics \textbf{5}, 499 (2011).

\bibitem{Mandatori} A. Mandatori, C Sibilia, M. Bertolotti, S. Zhukovsky,
J.W. Haus, and M. Scalora, Phys. Rev. B \textbf{70}, 165107 (2004).

\bibitem{Thaler} B. Thaller, arXiv:quant-ph/0409079 (2004).
\end{thebibliography}
\end{document}